\documentclass[times,review,preprint,authoryear]{elsarticle}

\usepackage{jasr}
\usepackage{framed,multirow}

\usepackage{amssymb}
\usepackage{latexsym}

\usepackage{amsmath}
\usepackage{amsfonts}
\usepackage{amsthm}
\usepackage{newlfont}
\usepackage{mathrsfs}
\usepackage{float}
\usepackage{ragged2e}
\usepackage{xcolor}
\usepackage{longtable}
\usepackage{multirow}
\usepackage{colortbl}
\usepackage{makecell}
\usepackage{caption}
\usepackage{subcaption}
\usepackage{lineno}

\usepackage{url}
\definecolor{newcolor}{rgb}{.8,.349,.1}

\usepackage[citebordercolor=white]{hyperref}

\journal{Advances in Space Research}

\newcommand{\ra}{\ddot{\mathbf{r}}}

\newcommand{\rhat}{\hat{\mathbf{r}}}
\newcommand{\nhat}{\hat{\mathbf{n}}}
\newcommand{\thetahat}{\hat{\boldsymbol{\theta}}}
\newcommand{\phihat}{\hat{\boldsymbol{\varphi}}}

\newcommand{\thetadot}{\dot{\theta}}

\newcommand{\phidot}{\dot{\varphi}}

\newcommand{\rdot}{\dot{r}}

\newcommand{\x}{\mathbf{x}}

\renewcommand{\d}{\mathrm{d}}

\begin{document}
	

\verso{Jeric Garrido \& Jose Perico Esguerra}

\begin{frontmatter}

\title{Trajectory design and optimization of a solar sail probe}

%

\author[1]{Jeric \snm{Garrido}\corref{cor1}}
\cortext[cor1]{Corresponding author} 
\ead{jgarrido@nip.upd.edu.ph}
\author[1]{Jose Perico  \snm{Esguerra}\fnref{fn1}}
\ead{jhesguerra1@up.edu.ph}

\address[1]{National Institute of Physics, University of the Philippines, Diliman, Quezon City 1101, Philippines}

\received{}
\finalform{}
\accepted{}
\availableonline{}
\communicated{}

\begin{abstract}

There is a desire to observe the sun's poles to further deepen our understanding of solar activity. However, because of the large speeds needed to perform out-of-ecliptic plane maneuvers, chemical and electric rocket propulsion mechanisms have been proven to be costly and impractical, leaving alternative space technology systems like solar sails to be considered for these applications. In this paper, we study the possibility of using a solar sail as a probe observing the sun. We design and optimize the trajectories of the solar sail probe through the surface constraint approach, with the assumption that the sail moves on a displaced spherical surface.  We first review the surface constraint approach, focusing on its important assumptions and limitations. Then, we solve and obtain a family of radial and azimuthal trajectory equations by choosing the correct constraint equation. We characterize the trajectories based on the functional dependence of the sail's orientation with the polar angle. Finally, we determine the trajectories of the probe that will give us the minimum flight time. Results show that increasing the number of mission stages decreases the total flight time, with minimal changes in the sail's radial and polar velocities. Furthermore, changing the functional dependence of the clock angle resets the azimuthal velocity, making the sail not reverse its direction and directly approach the sun along the spherical surface. 
\end{abstract}


\end{frontmatter}


\section{Introduction}
\label{sec:intro}

Solar sailing is a propulsion technology which uses the radiation pressure of the sun for space travel \citep{garrido2023surface,mcinnes2004solar,gong2019review}. Through the momentum transfer of solar  photons to its surface, a solar sail is capable of attaining its mission objectives without the need for a conventional chemical thruster. While this novel mission concept was  developed by Soviet scientists Tsiolkovsky and Tsander in the 1920s, it took almost a century for solar sailing to be successfully demonstrated with the launch of IKAROS in 2010 \citep{vulpetti2014solar,fu2016solar,mcinnes2004solar,tsuda2011flight}. Since then, additional sails like NanoSail D2, LightSail 1, and LightSail 2 have been successfully deployed, showing the feasibility of using solar sailing technology in various mission applications \citep{alhorn2011nanosail,betts2017lightsail,spencer2021lightsail}. The future of solar sailing is promising with the recently-launched NEA Scout mission by the NASA and the upcoming OKEANOS mission by JAXA \citep{pezent2018near,mori2020solar}, as well as other proposed missions like the Sundiver concept which may enable solar sails to reach their target destinations with ultrafast speeds \citep{turyshev2023science}.

Solar sails have potential applications in some missions where the use of conventional rocket propulsion mechanisms is impractical. For example, solar sails can be used in deep space exploration and missions towards Neptune and beyond \citep{vulpetti2012fast,lingam2020propulsion,dachwald2005optimal,macdonald2010technology}, in  generation of non-Keplerian orbits \citep{mckay2011survey,gong2011utilization,mcinnes1992solar,mcinnes1998dynamics}, and in planetary and asteroid observations \citep{lappas2009practical,ozimek2009design,bookless2008control,baoyin2006solar,waters2007periodic,peloni2016solar,song2019solar,morrow2001solar,macdonald2007geosail,zeng2015asteroid,zeng2016solar,zeng2019solar,heiligers2019trajectory}. For this paper though, we focus on one particular application which is on monitoring solar activity. There is a desire to study and perform measurements at the polar regions of the sun, which necessitates off-ecliptic plane orbits or trajectories with high inclination \citep{turyshev2023science}. Consequently, a large $\Delta v$ is necessary for these missions, which is costly and difficult for typical chemical and electric propulsion mechanisms \citep{liewer2013solar,turyshev2023science}. The continuously-applied nature of solar radiation force allows for these large changes in velocity, making it viable for these mission applications. For example, the GeoStorm mission concept was envisioned to provide real-time monitoring of Sun activity \citep{west2004geostorm,garrido2023surface}. On the other hand, the Solar Polar Imager (SPI) mission concept was proposed to place a light sail in a highly-inclined circular orbit of radius 0.48 AU and at least $0.75^{\circ}$ inclination to provide measurements from high altitudes \citep{thomas2020solar}. The proponents of the SPI mission claim that by observing the sun's poles, our understanding of the mechanism of solar activity cycles, polar magnetic field reversals, and the internal structure and dynamics of the sun will be revolutionized \citep{liewer2013solar}. The SPI mission concept has been referenced in other mission proposals, such as the High Inclination Solar Mission (HISM) concept which aims to observe the sun and the heliosphere using remote, \textit{in situ} and plasma wave instruments \citep{kobayashi2020high}. On the other hand, the POLARIS mission concept was proposed to address the limitation of the SPI mission, which is to determine the relationship between the magnetism and dynamics of the sun's polar regions \citep{appourchaux2009polar}. 

Two aspects of solar sailing research are trajectory design and optimization. Similar to any spacecraft in continuous propulsion, the trajectory equation of a solar sail can be obtained by solving its equation of motion, yielding numerical solutions. The nonlinear nature of the solar sail equation of motion makes obtaining optimal analytic trajectories difficult \citep{conway2010spacecraft}. However, for very special cases, semi-analytic solutions can also be determined, which can be used as preliminary guesses for more robust numerical solutions. A set of good initial guesses is necessary since the accuracy of the final trajectories mainly depend on these solutions. Some semi-analytic solutions for continuously-propelled systems such as the logarithmic spiral trajectory and the exponential sinusoid have been studied in Refs. \citep{petropoulos2002review,mckay2011survey}.

Different approaches have been developed to obtain these semi-analytic trajectories. For instance, the shape-based approach developed in \citet{petropoulos2003shape} and \citet{petropoulos2004shape} \textit{a priori} assumes a form of the trajectory equation with unknown parameter values. These parameters are then adjusted to match with the mission requirements. These shaping algorithms have been used in solving the Lambert's problem for exponential sinusoids \citep{izzo2006lambert}, as well as in three-dimensional rendezvous trajectories \citep{vasile2007optimality}. Other shaping methods include using Fourier series \citep{taheri2012shape} and Bezier curves \citep{fan2020initial} in trajectory design and optimization.

Previously, we proposed another approach of solar sail trajectory design and optimization, which is the surface constraint approach \citep{garrido2023surface}. Unlike shape-based algorithms which assume a form of the orbit or the trajectory equation, in this approach, we directly solve the solar sail equation of motion by deriving a generalized Laplace-Runge-Lenz vector, and then assume a surface where the sail is expected to move. The method was found to be useful in obtaining an optimal transfer towards the orbit of an asteroid with a highly-inclined orbit \citep{garrido2023surface,garrido2023flight}. 

In this paper, we explore the possibility of using a solar sail as a probe approaching the sun. To do this, we use the surface constraint approach in the design and flight time optimization of a solar sailing spacecraft for a sun probe mission. Unlike in our previous study where we applied the method to a solar sail on a cylindrical surface of revolution, we want to further show the method's versatility by solving a different surface, in this case a sphere displaced from the ecliptic plane. After obtaining a family of trajectories, we can then choose the one that gives us the minimum flight time.

The paper is organized as follows: We first review the important aspects of the surface constraint approach (Section \ref{sec:surface-constraint}). We then use this approach in obtaining the families of trajectories of a solar sail moving on a displaced spherical surface (Section \ref{sec:trajectories-displaced-sphere}). It turns out that the form of the trajectories depend on how the clock angle varies with the polar angle. Finally, we apply our approach in the design and optimization of a solar probe trajectory (Section \ref{sec:mission-application}).

\section{A Review of the Surface Constraint Approach}
\label{sec:surface-constraint}

We use the surface constraint approach to design the trajectories of a solar sail probe that is constrained to move on a displaced sphere \citep{garrido2023surface}. Let us briefly discuss this approach in this section. 

\begin{figure}[H]
	\centering
	\includegraphics[scale=0.55]{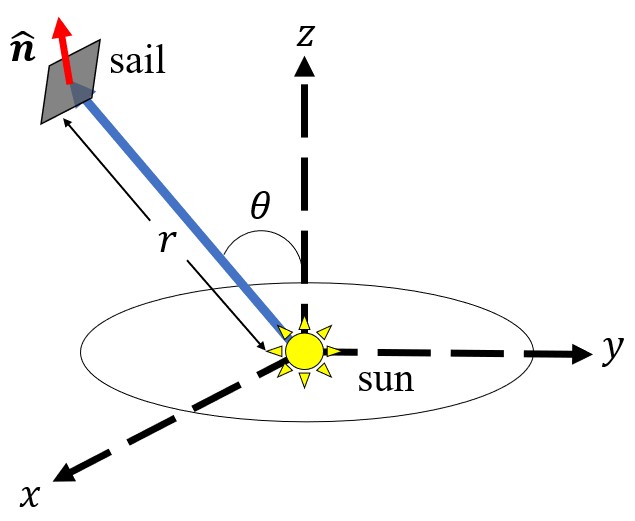}
	\caption{A solar sail on a three-dimensional heliocentric inertial reference frame.}
	\label{fig:system-considered}
\end{figure}

Consider a solar sail that is in a heliocentric inertial reference frame represented using spherical coordinates $(r,\theta,\varphi)$. This sail follows ideal force model such that it is rigid and flat, its surface is perfectly-reflecting, and the sun is an ideal blackbody and a point source of radiation. Then, if $\nhat$ is the unit normal to the sail's surface, the equation of motion in coordinate-free form, using a heliocentric inertial reference frame, is given by
\begin{equation}
	\ra=-\frac{\mu}{r^{2}}\rhat+\frac{\beta\mu}{r^{2}}(\nhat\cdot\rhat)^{2}\rhat
\end{equation}
where $\beta$ is the sail's lightness number or the ratio of the solar radiation pressure force and the gravitational force, $\mu=GM_{\odot}$ is the universal gravitational parameter, $G$ is the universal gravitational constant and $M_{\odot}$ is the solar mass. We solve this equation of motion, yielding solutions of the form $r(\theta,\varphi)$. \textit{(See Figure \ref{fig:system-considered})}
Representing the unit normal vector in this coordinate system, two angles define the unit vector: the cone angle which measures the angle between the radial unit vector $\rhat$ and $\nhat$, and the clock angle which measures the projection of $\nhat$ with the local horizontal plane \citep{garrido2023surface}. Consequently, the unit normal vector can be decomposed into its radial, azimuthal and polar components as follows:
\begin{equation}
	\begin{split}
		\nhat&=\rhat\cos\alpha+\thetahat\sin\alpha\cos\delta+\phihat\sin\alpha\sin\delta\\
		&= \rhat n_{r}+\thetahat n_{\theta}+\phihat n_{\varphi}
	\end{split}
\end{equation}

We have shown in \citet{garrido2023surface,garrido2023flight} that if the radial and polar components of the velocity are related by a function $G(\theta)$, which we define as the constraint equation, that is, if
\begin{equation}
	\label{eq:constraint_eqn}
	\rdot=r G(\theta)\thetadot,
\end{equation}
then one can determine a generalized Laplace-Runge-Lenz (LRL) vector, that in return can be used to generate orbit and/or trajectory equations. This is possible provided that the cone angle, and hence the radial component of the unit normal vector, is constant throughout the mission. However, there is no limit on the form of the polar and azimuthal components, except the fact that the unit vector must be normalized.  

Consequently, the trajectory of the sail can be specified using two equations: the radial equation which relates the radial and polar components, and the azimuthal equation which relates the azimuthal and polar coordinates. In our approach, the polar angle $\theta$ is the independent parameter. If $r_{0}$ is the initial position of the sail, $J$ is the squared magnitude of the LRL vector, then the radial equation takes the form
\begin{equation}
	\label{eq:radial_eqn}
	r(\theta)=r_{0}\exp\left(\int_{\theta_{0}}^{\theta}\d\nu\ G(\nu)\right)=\frac{h^{2}}{\mu+J-\beta\mu n_{r}^{2}}\exp\left(\int_{\theta_{0}}^{\theta}\d\nu\ G(\nu)\right).
\end{equation}
On the other hand, the azimuthal equation is given by
\begin{equation}
	\label{eq:azm_eqn}
	\varphi(\theta)=\varphi(\theta_{0})+\int_{\theta_{0}}^{\theta}\d\nu \frac{2n_{\varphi}(\nu)+A(\nu)}{\sin\nu(2n_{\theta}(\nu)+B(\nu))}
\end{equation}
where
\begin{equation}
	A(\nu)=\int_{\theta_{0}}^{\nu}\d\eta\ n_{\varphi}(\eta) G(\eta)
\end{equation}
and
\begin{equation}
	B(\nu)=\int_{\theta}^{\nu}\d\eta\ n_{\theta}(\eta)G(\eta)\cos(\nu-\eta).
\end{equation}

These two equations define the sail's trajectory in space. The radial equation specifies the surface where the sail is expected to move \citep{garrido2023surface}. While the radial equation is given \textit{a priori} by the constraint equation, the surface constraint approach assures that the trajectory of the sail satisfies the solar sail equation of motion. In fact, it was shown that for the solution not to be extraneous, the following inequality, known as the complete trajectory characterization condition, must be satisfied:
\begin{equation}
	\label{eq:complete_char}
	J=-\mu+\beta\mu n_{r}^{3}+\frac{2\beta\mu}{G(\theta_{0})}\left(\frac{n_{r}^{2}(1-n_{r}^{2})}{n_{\theta}(\theta_{0})}\right)\ge 0.
\end{equation}

On the other hand, the azimuthal equation specifies the path taken by the light sail in space. While the radial equation mainly depends on the cone angle, the azimuthal equation depends only on the sail's clock angle, a consequence of the condition that the cone angle is always constant throughout the mission. In return, the azimuthal equation is scale-independent relative to the lightness number. 

Other physical quantities from the constraint, radial and azimuthal equations. For example, the time derivatives $\rdot$, $\thetadot$ and $\phidot$ can be determined directly from equations \eqref{eq:constraint_eqn}, \eqref{eq:radial_eqn} and \eqref{eq:azm_eqn}:
\begin{equation}
	\label{eq:thetadot}
	\displaystyle{\thetadot=\pm \left[\frac{\beta\mu n_{r}^{2}}{r^{3}G(\theta)}\left(2n_{\theta}(\theta)+\int_{\theta_{0}}^{\theta}\d\eta\ n_{\theta}(\eta)G(\eta)\cos(\theta-\eta)\right)\right]^{1/2}}.
\end{equation}
\begin{equation}
	\label{eq:rdot}
	\rdot=\pm \left(\frac{\beta\mu}{r}n_{r}^{2}G(\theta)\left[2n_{\theta}(\theta)+\int_{\theta_{0}}^{\theta}\d\eta\ n_{\theta}(\eta)G(\eta)\cos(\theta-\eta)\right]\right)^{1/2},
\end{equation}
and
\begin{equation}
	\label{eq:phidot}
	\phidot=\pm \frac{\d\varphi}{\d\theta}\frac{\d\theta}{\d t}=\pm \left[\frac{\beta\mu n_{r}^{2}}{r^{3}G(\theta)}\right]^{1/2}\frac{2n_{\varphi}(\theta)+\int_{\theta_{0}}^{\theta}\d\eta \ n_{\varphi}(\eta)G(\eta)}{\sin\theta\left[2n_{\theta}(\theta)+\int_{\theta_{0}}^{\theta}\d\eta\ n_{\theta}(\eta)G(\eta)\cos(\theta-\eta)\right]^{1/2}}.
\end{equation}

Finally, from the polar time derivative $\thetadot$, we can also determine the total flight time of the sail from its initial position $(r_{0},\theta_{0},\varphi_{0})$ to its final position $(r_{f},\theta_{f},\varphi_{f})$:
\begin{equation}
	\begin{split}
		\label{eq:time-of-flight}
		T(\theta_{0}|\theta_{f})=\displaystyle{\int_{\theta_{0}}^{\theta_{f}}\d\theta\left(\frac{r^{3}G(\theta)/\beta\mu n_{r}^{2}}{2n_{\theta}(\theta)+\int_{\theta_{0}}^{\theta}\d\eta\  n_{\theta}(\eta)\ G(\eta) \cos(\theta-\eta)}\right)^{1/2}}.
	\end{split}
\end{equation}

In summary, the surface constraint approach consists of the following steps:
\begin{itemize}
	\item Determine the most appropriate geometry of the surface and the form of $n_{\theta}$ (or $n_{\varphi}$) based on the mission requirements
	\item From the chosen surface geometry, choose the correct surface constraint equation \eqref{eq:constraint_eqn}.
	\item Equation \eqref{eq:constraint_eqn} immediately gives us the correct radial and azimuthal equations depending on the initial position $(r_{0},\theta_{0},\varphi_{0})$ and velocity $(\rdot_{0},\thetadot_{0},\phidot_{0})$. 
	\item From the radial and azimuthal equations, other physical quantities such as velocities and flight time can then be determined. 
\end{itemize}

\section{Trajectories on a Displaced Sphere}
\label{sec:trajectories-displaced-sphere}

\subsection{Constraint Equation}
As an application of the surface constraint approach, let us consider a light sail that is constrained to move on a sphere of radius $R$ and is centered at $(0,0,a)$ of our heliocentric coordinate system. Note that $a$ here can be positive or negative, where $a>0$ ($a<0$) means that the center of the displaced sphere is above (below) the ecliptic plane. 
\begin{figure}[H]
	\centering
	\includegraphics[scale=0.45]{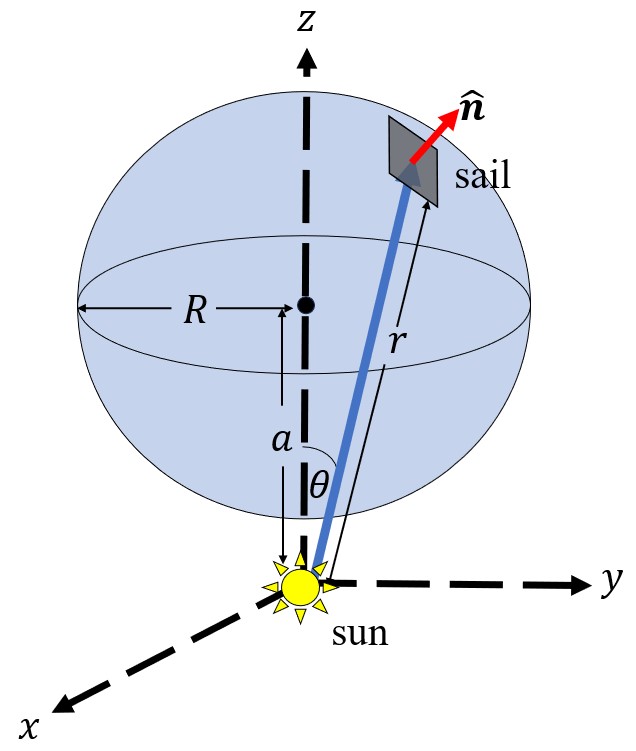}
	\caption{A sail's trajectory constrained on a displaced sphere. The shaded region is the surface where the sail is constrained to move.}
	\label{fig:cyl_diagram}
\end{figure}
We first obtain the constraint equation appropriate for this system. In rectangular coordinates, the equation of the sphere is given by 
\begin{equation}
	x^{2}+y^{2}+(z-a)^{2}=R^{2}.
\end{equation}
This can be re-written in spherical coordinates using the transformation $x=r\sin\theta\cos\varphi$, $y=r\sin\theta\sin\varphi$, and $z=r\cos\theta$. The resulting equation is then
\begin{equation}
	\label{eq:radial_eqn_dsp_sph}
	r=a\cos\theta\pm \sqrt{R^{2}-a^{2}\sin^{2}\theta}.
\end{equation}
To determine $G(\theta)$, we differentiate $r$ with respect to $\theta$, giving us
\begin{equation}
	\frac{\d r}{\d\theta}=\mp \frac{a\sin\theta}{\sqrt{R^{2}-a^{2}\sin^{2}\theta}}\left(a\cos\theta\pm \sqrt{R^{2}-a^{2}\sin^{2}\theta}\right)=\mp r\frac{a\sin\theta}{\sqrt{R^{2}-a^{2}\sin^{2}\theta}}
\end{equation}
Comparing this equation with $\frac{\d r}{\d\theta}=rG(\theta)$, the constraint equation is thus given by
\begin{equation}
	G(\theta)=\mp \frac{a\sin\theta}{\sqrt{R^{2}-a^{2}\sin^{2}\theta}}.
\end{equation}
The positive root takes the upper half of the sphere from $\theta\in [0,\pi/2)$ while the negative root takes the lower half of the spherical surface \textit{i.e.} $\theta\in [\pi/2,\pi)$. While in this paper, we focus on the positive root, it is possible to generate trajectories using the negative root of the radial equation.  Using the positive root, the radial equation takes the form
\begin{equation}
	r(\theta)=a\cos\theta+ \sqrt{R^{2}-a^{2}\sin^{2}\theta}
\end{equation}
while the azimuthal equation depends on the functional dependence of the clock angle on $\theta$. 

\subsection{Constant Clock Angles}
The path of the sail depends mainly on the choice of $\delta(\theta)$. Let us consider the case when both the cone and clock angles are constant throughout the mission such that $\delta(\theta)=\delta_{0}$. Then, the azimuthal equation is given by
\begin{equation}
	\label{eq:dsp_sph_azm_orb_eqn_constant}
	\varphi(\theta)=\varphi(\theta_{0})+\tan\delta_{0}\int_{\theta_{0}}^{\theta}\d\nu\dfrac{2-\int_{\theta_{0}}^{\nu}\d\eta\ \dfrac{a\sin\eta}{\sqrt{R^{2}-a^{2}\sin^{2}\eta}} }{\sin\nu \left[2-\int_{\theta_{0}}^{\nu}\d\eta\ \dfrac{a\cos(\eta-\nu)\sin\eta}{\sqrt{R^{2}-a^{2}\sin^{2}\eta}}\right]}.
\end{equation}
The above equation can be evaluated numerically, which together with the radial equation, gives us the trajectory of the sail. 

\begin{figure*}[h!]
	\centering
	\begin{subfigure}{0.55\textwidth}
		\centering
		\includegraphics[scale=0.18]{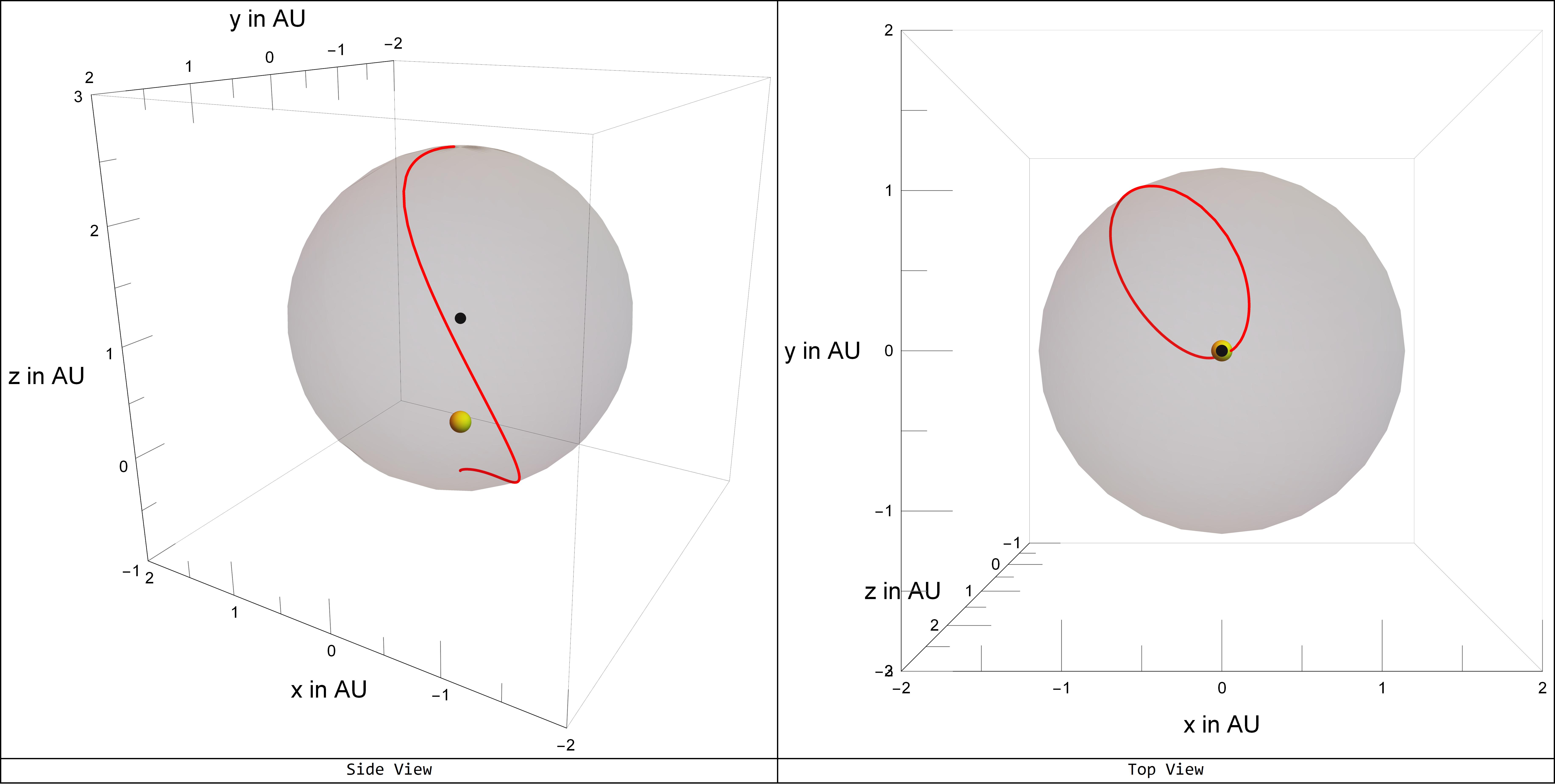}
		\caption{$\delta=30^{\circ}$}
		\label{fig:delta_30}
	\end{subfigure}%
	\begin{subfigure}{0.55\textwidth}
		\centering
		\includegraphics[scale=0.18]{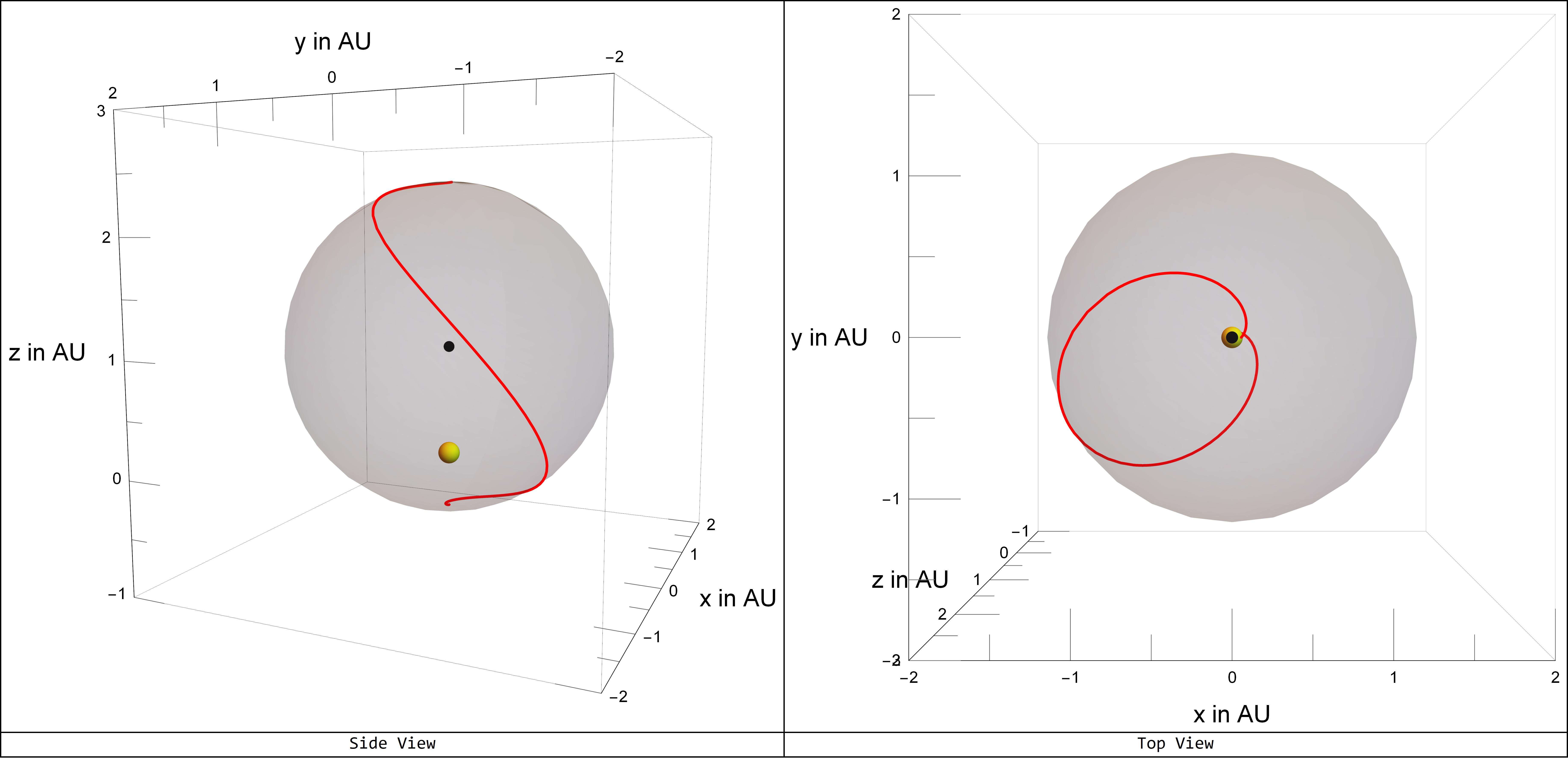}
		\caption{$\delta=45^{\circ}$}
		\label{fig:delta_45}
	\end{subfigure}
	\begin{subfigure}{0.55\textwidth}
		\centering
		\includegraphics[scale=0.18]{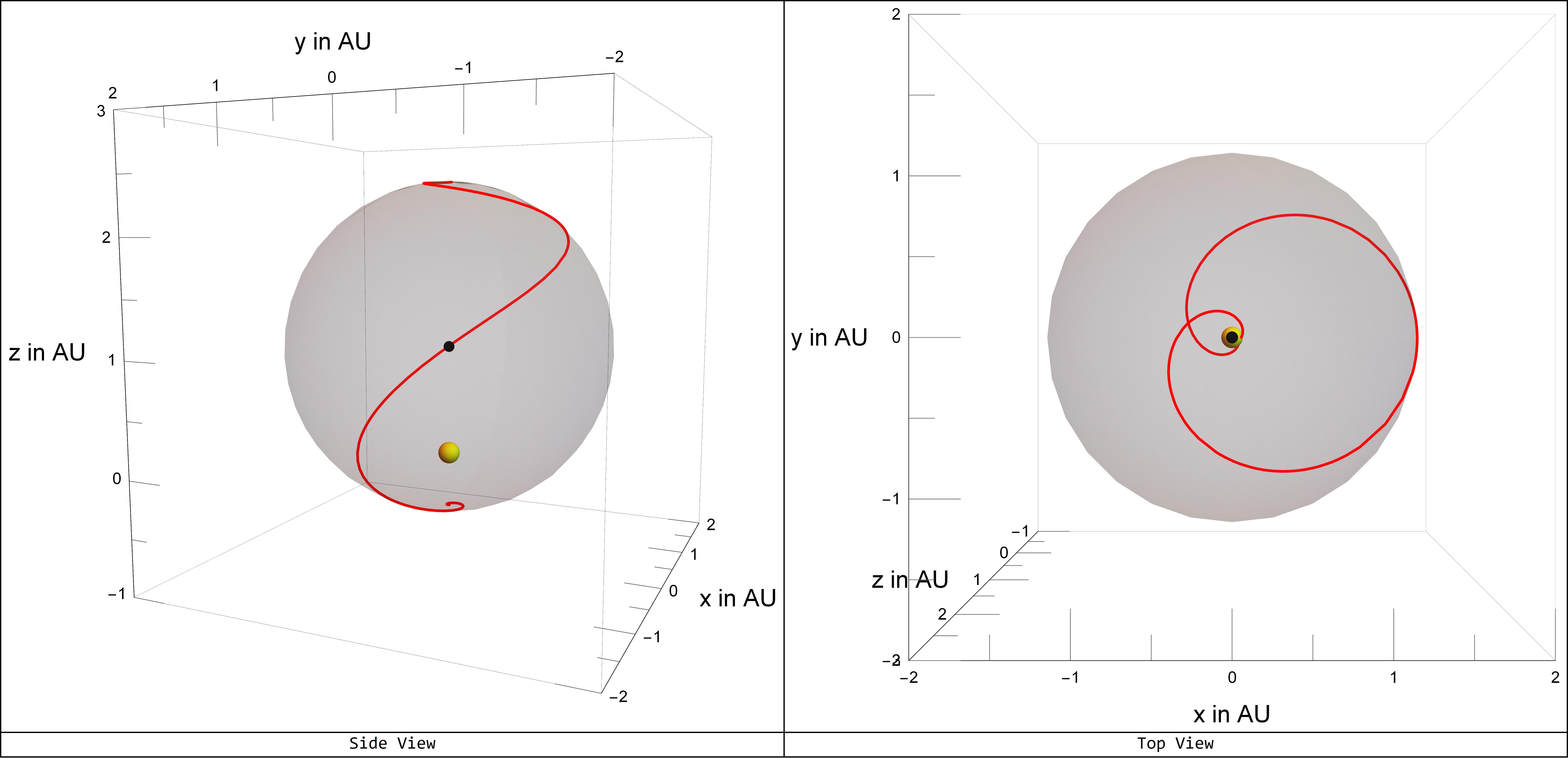}
		\caption{$\delta=60^{\circ}$}
		\label{fig:delta_60}
	\end{subfigure}%
	\begin{subfigure}{0.55\textwidth}
		\centering
		\includegraphics[scale=0.18]{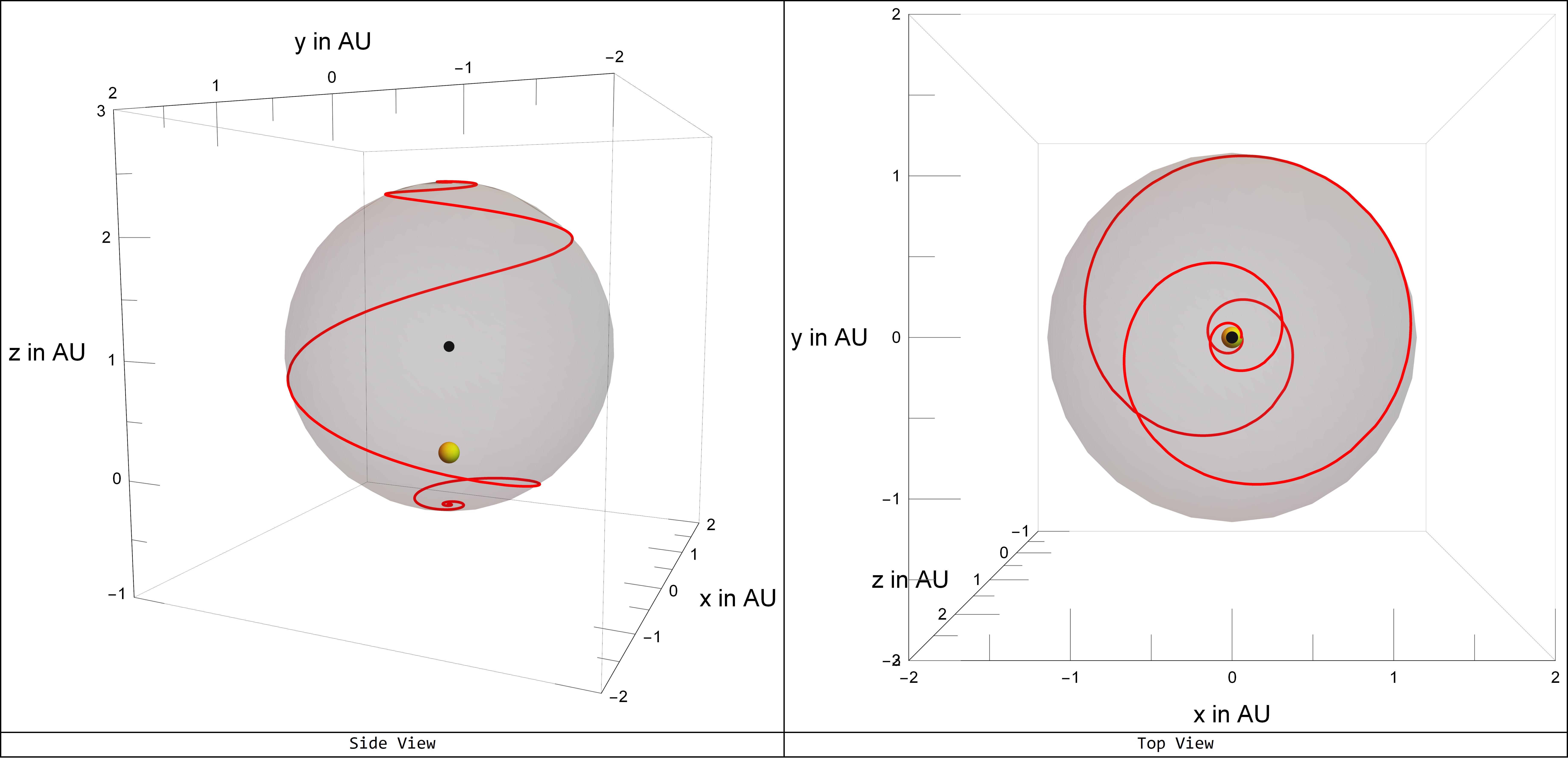}
		\caption{$\delta=75^{\circ}$}
		\label{fig:delta_75}
	\end{subfigure}
	\caption{Trajectories constrained on a displaced sphere for different constant clock angles.}
	\label{fig:dsp_sph_constant_orbits}
\end{figure*}

Figure \ref{fig:dsp_sph_constant_orbits} shows the trajectories of the sail for different values of the clock angle $\delta_{0}$. In our simulations, we assume that $a=1.00$ AU, $R=1.5$ AU, and the cone angle is at $-35.26^{\circ}$, which is the optimal value for $\alpha$. As the clock angle increases, there is also an increase in the number of turns in the trajectory. This is signified by the large azimuthal displacement between the sail at $\theta=0^{\circ}$ and at $\theta=180^{\circ}$.

These results are consistent with what is shown in \citet{garrido2023surface}, which is unsurprising because in both cases, the azimuthal displacement varies linearly with $\tan\delta_{0}$. At a specific value of $\theta$ the integral reduces to a constant, and thus is independent of the form of $\delta_{0}$. In fact. one can show that this is true for any non-trivial form of $G(\theta)$. That is, if the clock angle is constant with respect to the polar angle, then
\begin{equation}
	\varphi(\theta)=\varphi(\theta_{0})+(\tan\delta_{0})\tilde{\varphi}(G,\theta).
\end{equation}
where $\tilde{\varphi}(G,\theta)$ is the integral term in Eq. \eqref{eq:azm_eqn}. 

\subsection{Periodic $n_{\theta}$ and $n_{\varphi}$}
\begin{figure*}[h!]
	\centering
	\begin{subfigure}{0.55\textwidth}
		\centering
		\includegraphics[scale=0.18]{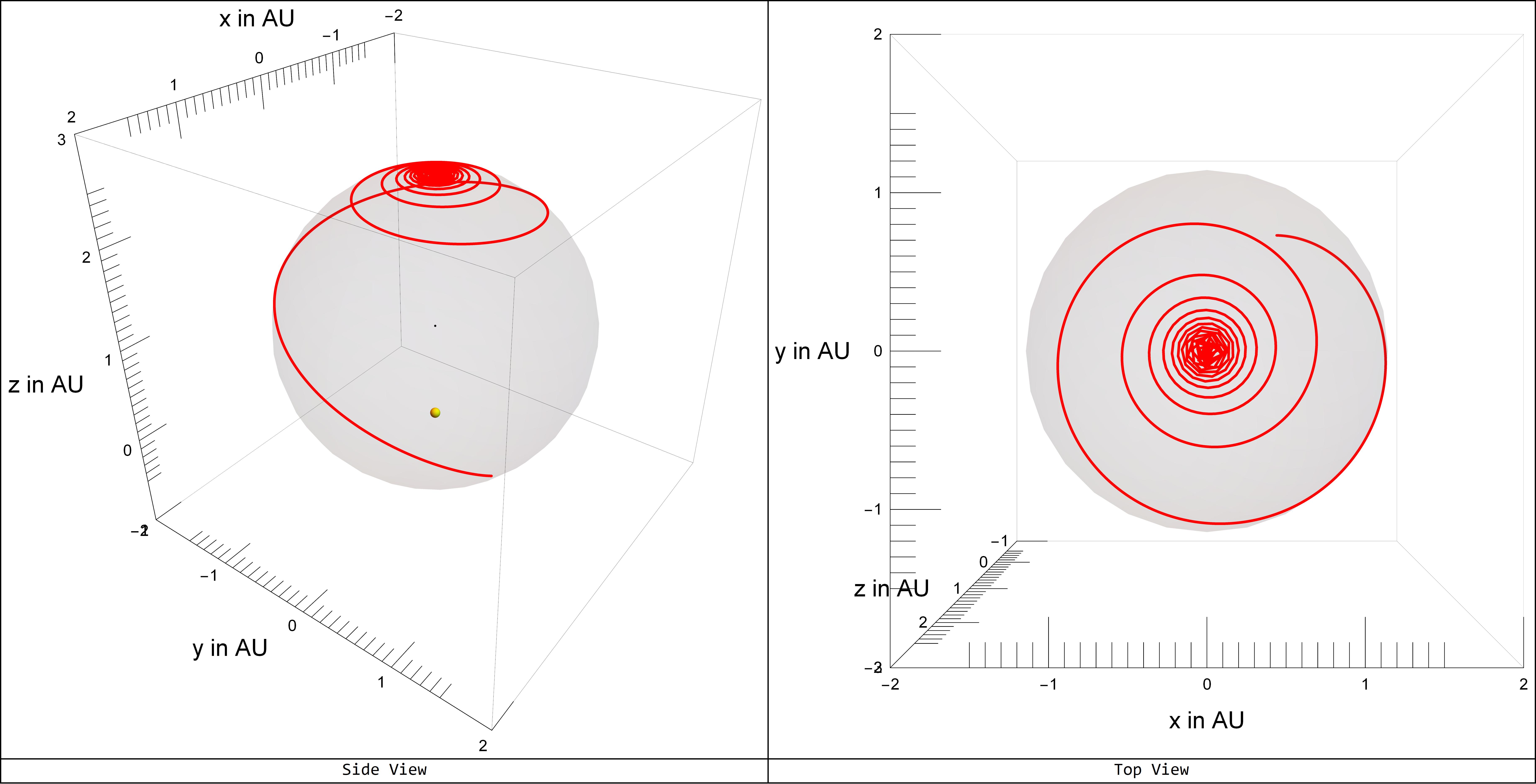}
		\caption{$k=1/3$}
		\label{fig:k_1-3}
	\end{subfigure}%
	\begin{subfigure}{0.55\textwidth}
		\centering
		\includegraphics[scale=0.18]{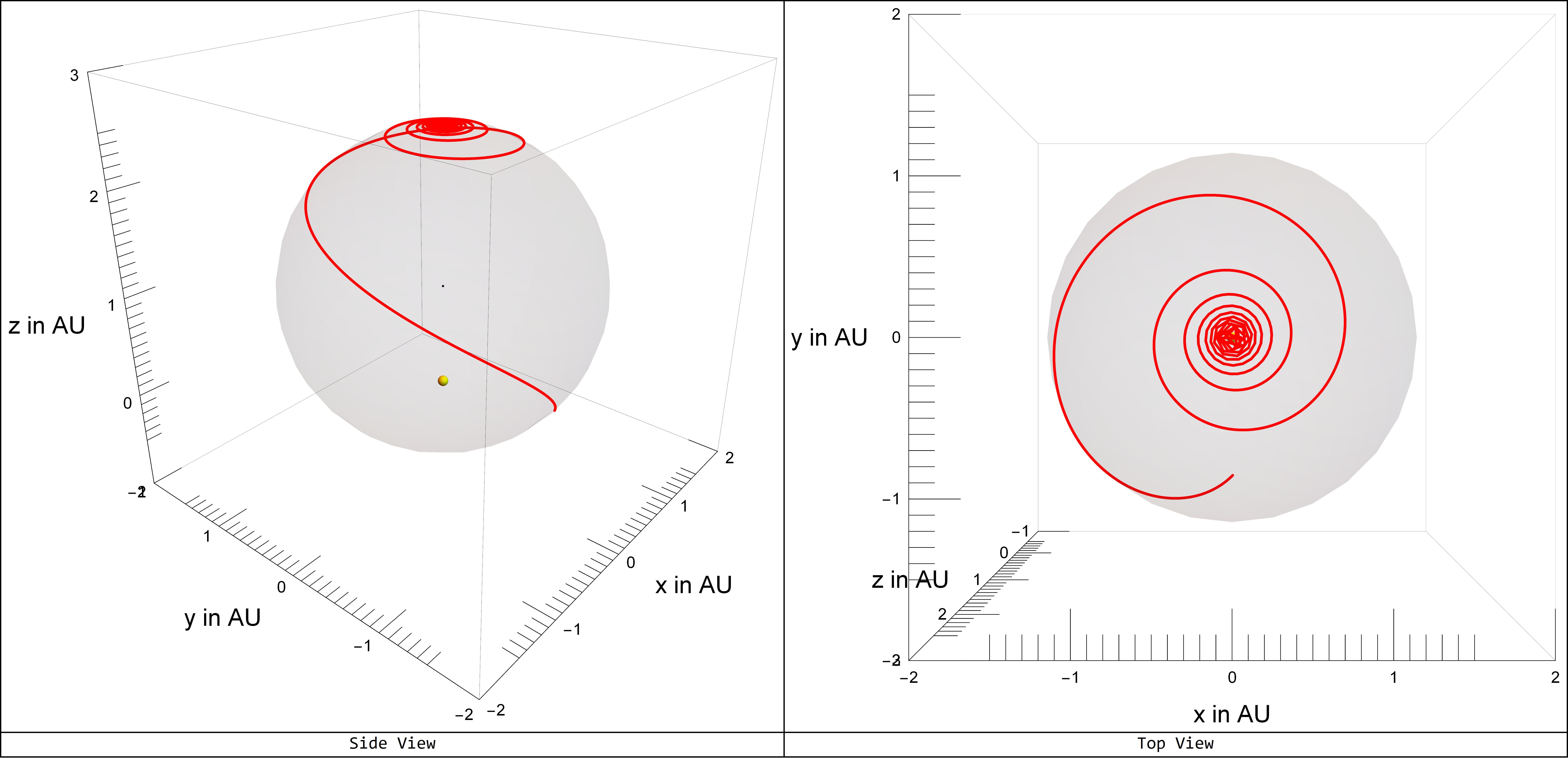}
		\caption{$k=1/2$}
		\label{fig:k_1-2}
	\end{subfigure}
	\begin{subfigure}{1\textwidth}
		\centering
		\includegraphics[scale=0.18]{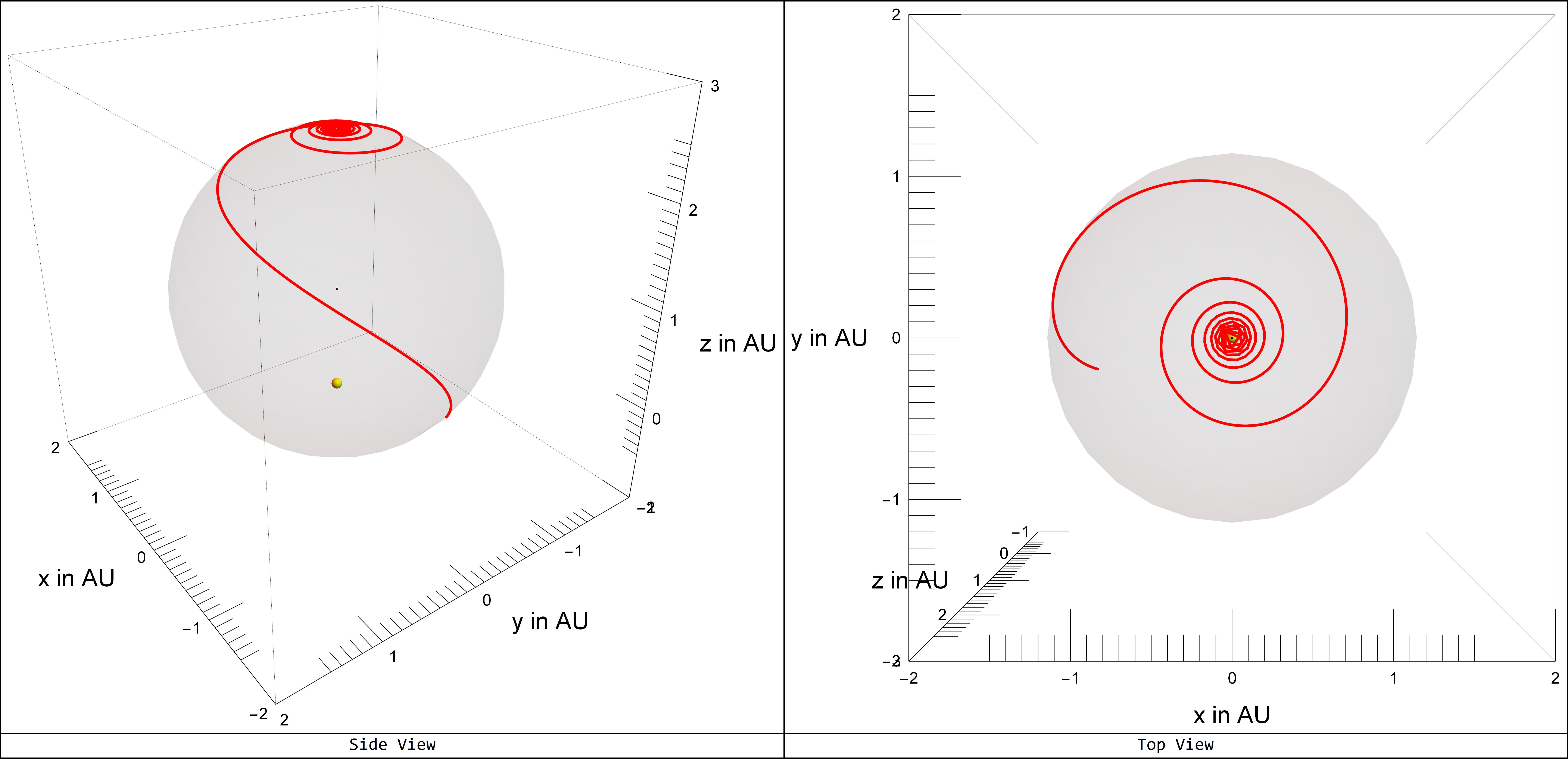}
		\caption{$k=2/3$}
		\label{fig:k_2-3}
	\end{subfigure}
	\begin{subfigure}{0.55\textwidth}
		\centering
		\includegraphics[scale=0.18]{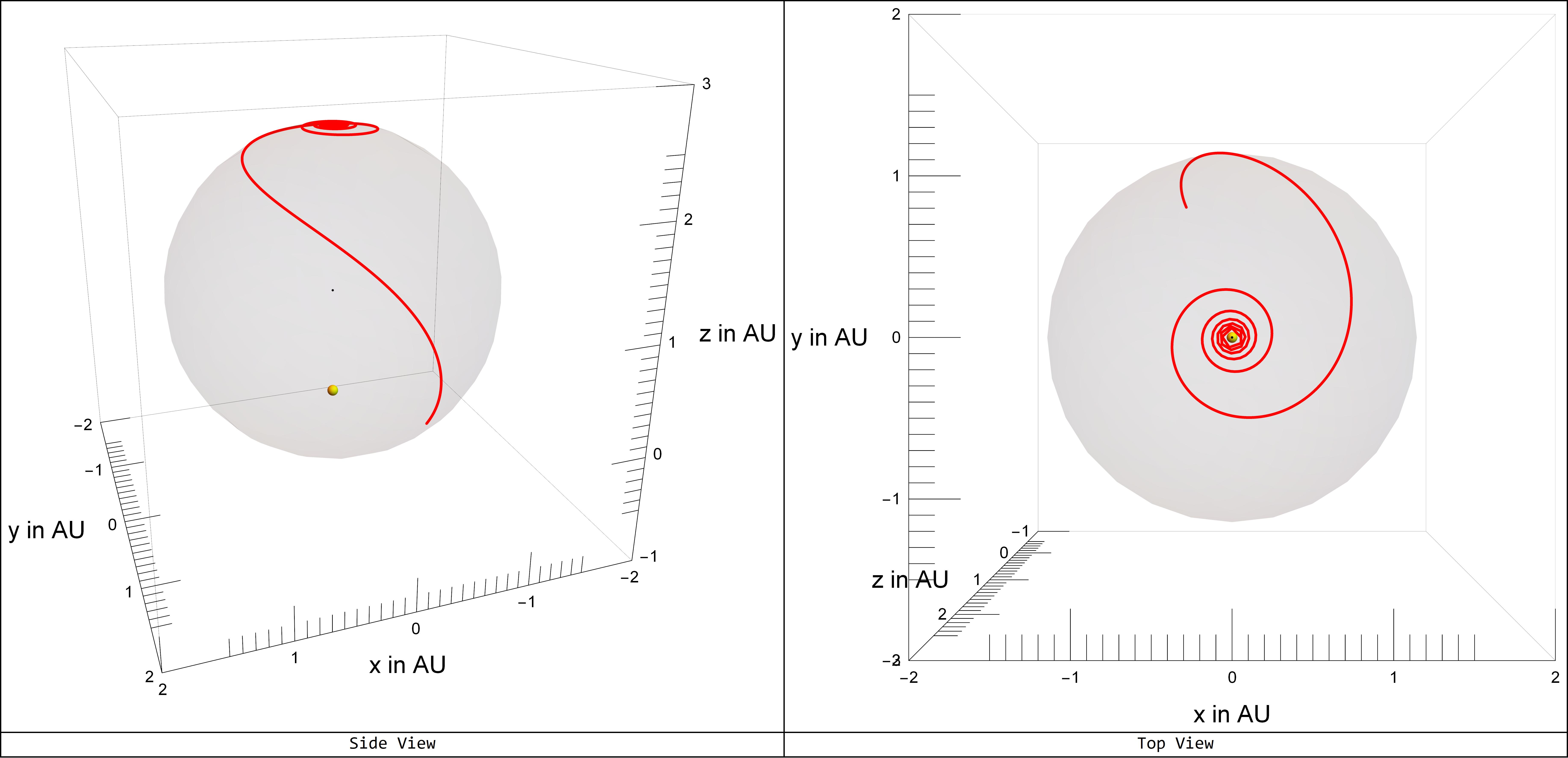}
		\caption{$k=1$}
		\label{fig:k_1}
	\end{subfigure}%
	\begin{subfigure}{0.55\textwidth}
		\centering
		\includegraphics[scale=0.18]{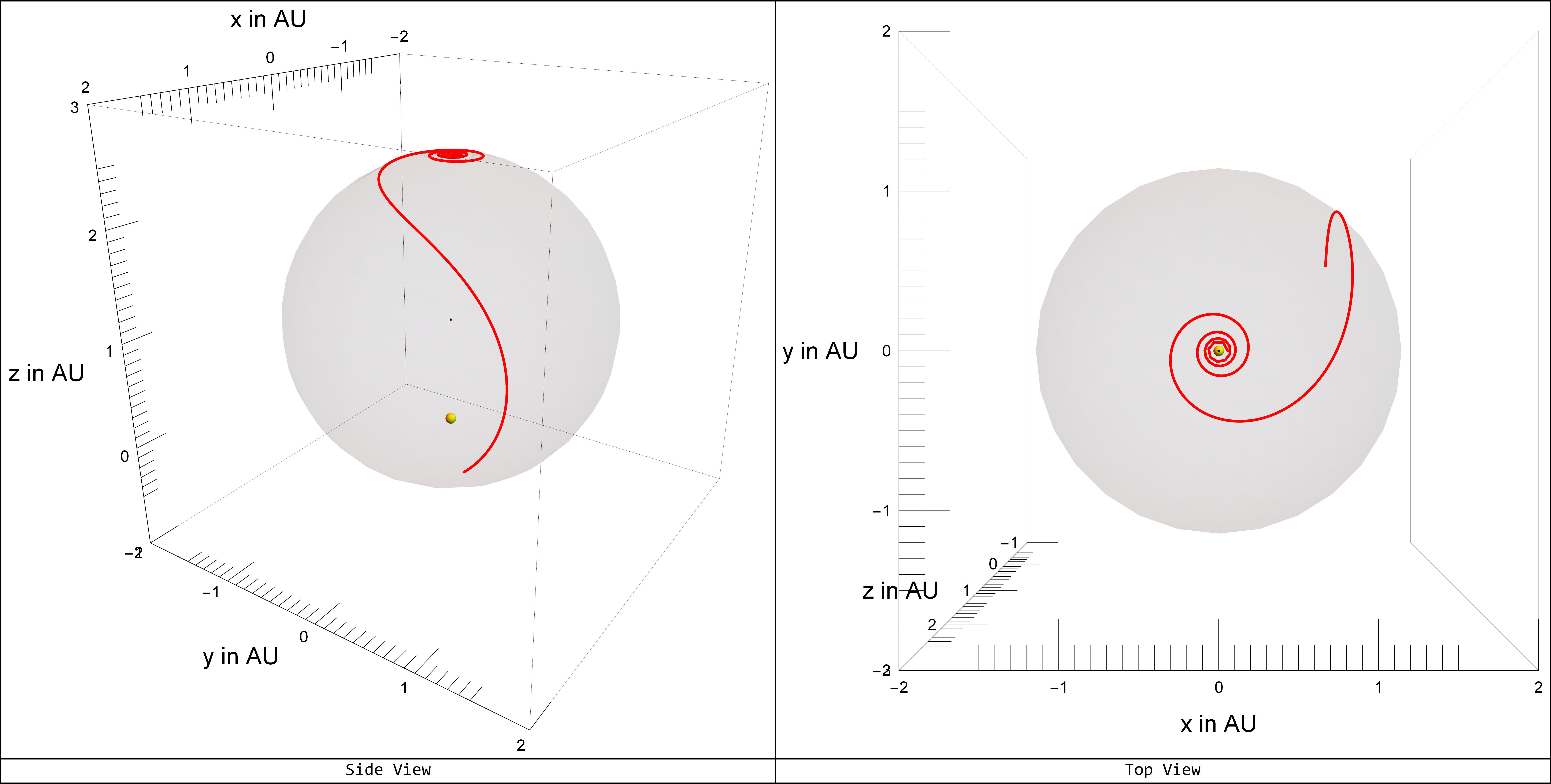}
		\caption{$k=3/2$}
		\label{fig:k_3-2}
	\end{subfigure}
	\begin{subfigure}{1\textwidth}
		\centering
		\includegraphics[scale=0.18]{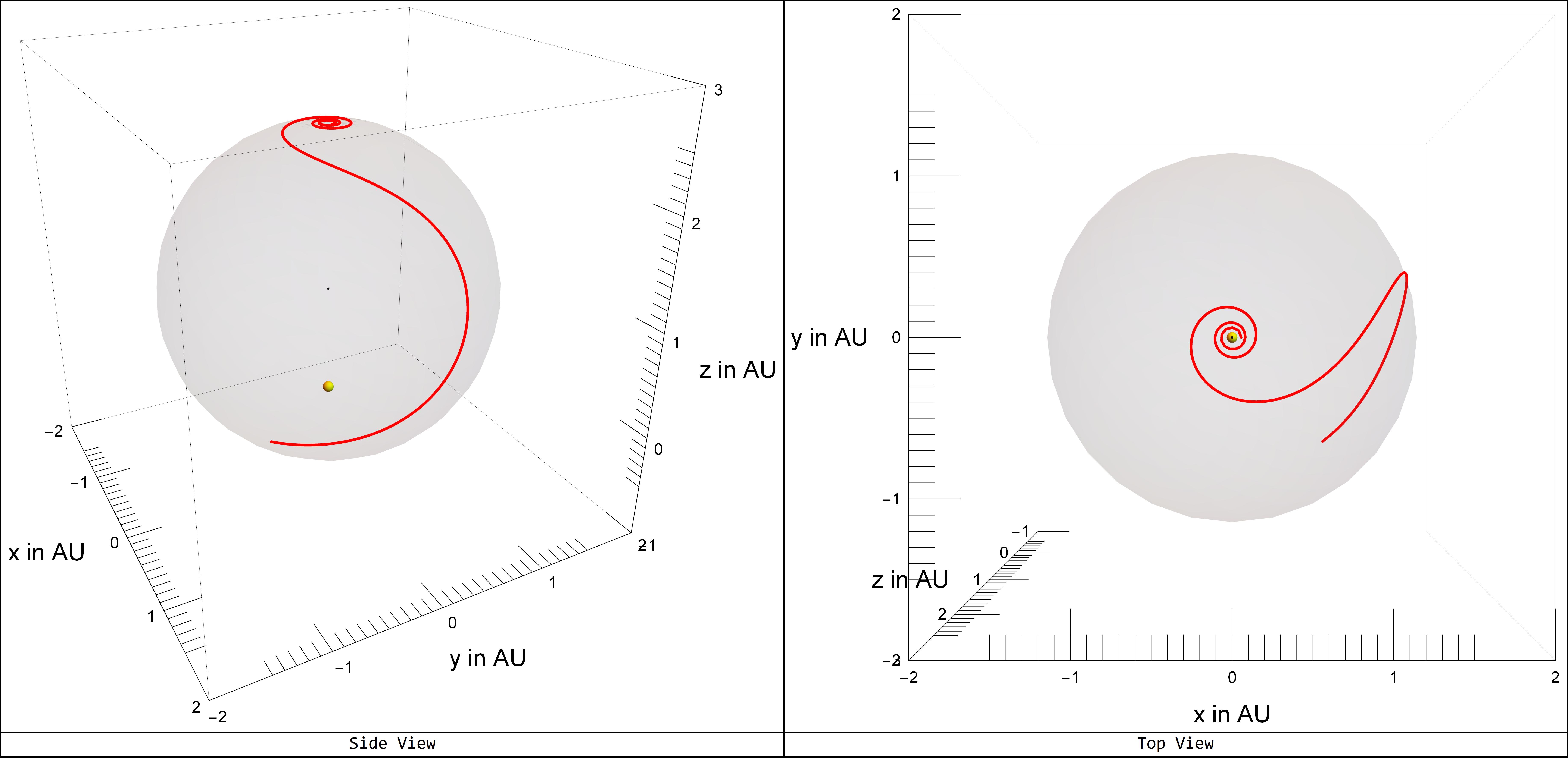}
		\caption{$k=2$}
		\label{fig:k_2}
	\end{subfigure}
	\caption{Trajectories constrained on a displaced sphere for different values of $k$.}
	\label{fig:dsp_sph_periodic_orbits}
\end{figure*}

It is possible to use a different form of the clock angle, as long as the normalization condition for $\nhat$ is satisfied. Let us consider the following functional dependence of $n_{\theta}$ and $n_{\varphi}$, for some constant $k$:
\begin{equation}
	\label{eq:periodic_ntheta_nphi}
	n_{\theta} = \sqrt{1-n_{r}^{2}}\sin k\theta; \quad \quad n_{\varphi}=\sqrt{1-n_{r}^{2}}\cos k\theta
\end{equation}
The constant $\sqrt{1-n_{r}^{2}}$ assures that the unit vector is normalized. Then, the azimuthal orbit equation takes the form
\begin{equation}
	\varphi(\theta)=\varphi(\theta_{0})+\int_{\theta_{0}}^{\theta}\d\nu \dfrac{2\cos k\nu +\int_{\theta_{0}}^{\nu}\d\eta\ \dfrac{a\cos k\eta\sin\eta}{\sqrt{R^{2}-a^{2}\sin^{2}\eta}}}{\sin\nu\left(2\sin k\nu +\int_{\theta}^{\nu}\d\eta\ \dfrac{a\sin k\eta \sin\eta \cos(\eta-\nu)}{\sqrt{R^{2}-a^{2}\sin^{2}\eta}}\right)}.
\end{equation}
The trajectories are shown in Figure \ref{fig:dsp_sph_periodic_orbits} for different values of $k$. The trajectories exhibit different characteristics for $k<1$ and $k\ge 1$. For $k$ values smaller than 1 shown in Figures \ref{fig:k_1-3} to \ref{fig:k_2-3}, the azimuthal position of the sail increases monotonically as $\theta$ increases. As $k$ decreases, the total number of revolutions increases. This is due to the relationship between $\delta$ and $\theta$. For smaller values of $k$, it will take a larger value of $\theta$ for $\delta $ to reach $90^{\circ}$, hence we see larger azimuthal displacements, and more revolutions for these values of $k$. 

An interesting trend occurs when $k\ge 1$. For this set of $k$ values, the clock angle will reach $90^{\circ}$ even with a polar angle smaller than $90^{\circ}$, changing the sign of $n_{\theta}$. Consequently, the azimuthal displacement becomes negative for angles greater than $90^{\circ}$, reversing the trajectory in the azimuthal direction. This reversal is evident in the trajectories shown in Figures \ref{fig:k_1} to \ref{fig:k_2}. 

By definition, $\delta$ only takes values from $-180^{\circ}$ to $180^{\circ}$. Hence, for values of $k$ larger than 1, there is an critical value of $\theta$ for which the upper limit of the clock angle is reached. Beyond this value, the trajectory becomes ill-defined so that a re-definition of the clock angle is necessary. We note that this upper limit of $\delta$ is reached at about $\theta=82.5^{\circ}$ for $k=2$.

Finally, we compare our results to those written in the literature. The increasing number of revolutions for $0<k<1$, and the apparent reversal of the trajectories for larger values of $k$ are also observed in previous cases, such as trajectories that are constrained on a cylinder, which we previously studied in \citet{garrido2023surface}. However, one striking difference between trajectories on a displaced sphere and on a cylinder is on the behavior near $\theta=0$. For trajectories constrained on a displaced sphere, more revolutions are observed near the ecliptic pole than near the ecliptic plane \citep{garrido2023surface}. This is different from trajectories constrained on a cylinder where the paths cluster near $\theta=90^{\circ}$.

%

\section{Mission application: a solar probe}
\label{sec:mission-application}

A possible mission application of orbits constrained on displaced spheres is on observing solar activity. It is possible to use the surface constraint approach to determine the optimal transfer towards a terminal location close to the sun, making the sail a solar probe. We consider the range of distances when the sail still follows nearly ideal behavior and its film's structure stays intact and does not deteriorate as the sail approaches the sun. Consequently, we consider the case when the light sail's terminal position is at $\theta=80^{\circ}$. 

We first assume that the sail is located at a distance of $1 \text{ AU}$ above the ecliptic plane of the sun. Then, the sail has a lightness number of $\beta=0.15$ and follows the radial equation \eqref{eq:radial_eqn_dsp_sph}, where $R=a=0.5\text{ AU}$. Further, we consider the case when the polar and azimuthal components are periodic in $\theta$ such that the sail's control follows equation \eqref{eq:periodic_ntheta_nphi}. 

We determine the trajectories that will make the flight time a minimum by splitting the trajectories into different sub-intervals such that each stage has a unique value of $k$. Assuming that the control switch is done instantaneously, for $n$ sub-intervals the total flight time is given by
\begin{equation}
	T(\theta_{0}|\theta_{f}) =\sum_{i=0}^{N}T_{k_{i}}(\theta_{i}|\theta_{i+1}).
\end{equation}
For the single stage trajectory, we determine the correct value of $k$ that will make the flight time a minimum. For piecewise continuous trajectories, we determine the transition points and the value of $k$ in each sub-trajectory. We consider a single optimal trajectory, a trajectory with two stages, a trajectory with three stages, and a trajectory with four stages in this paper. We then use Brent's method for the single-stage trajectory optimization while we use Trust Region constrained optimization for other three trajectories \citep{brent2013algorithms,forsythe1977computer,conn2000trust}.

\begin{figure*}[h!]
	\centering
	\begin{subfigure}{1\textwidth}
		\centering
		\includegraphics[scale=0.32]{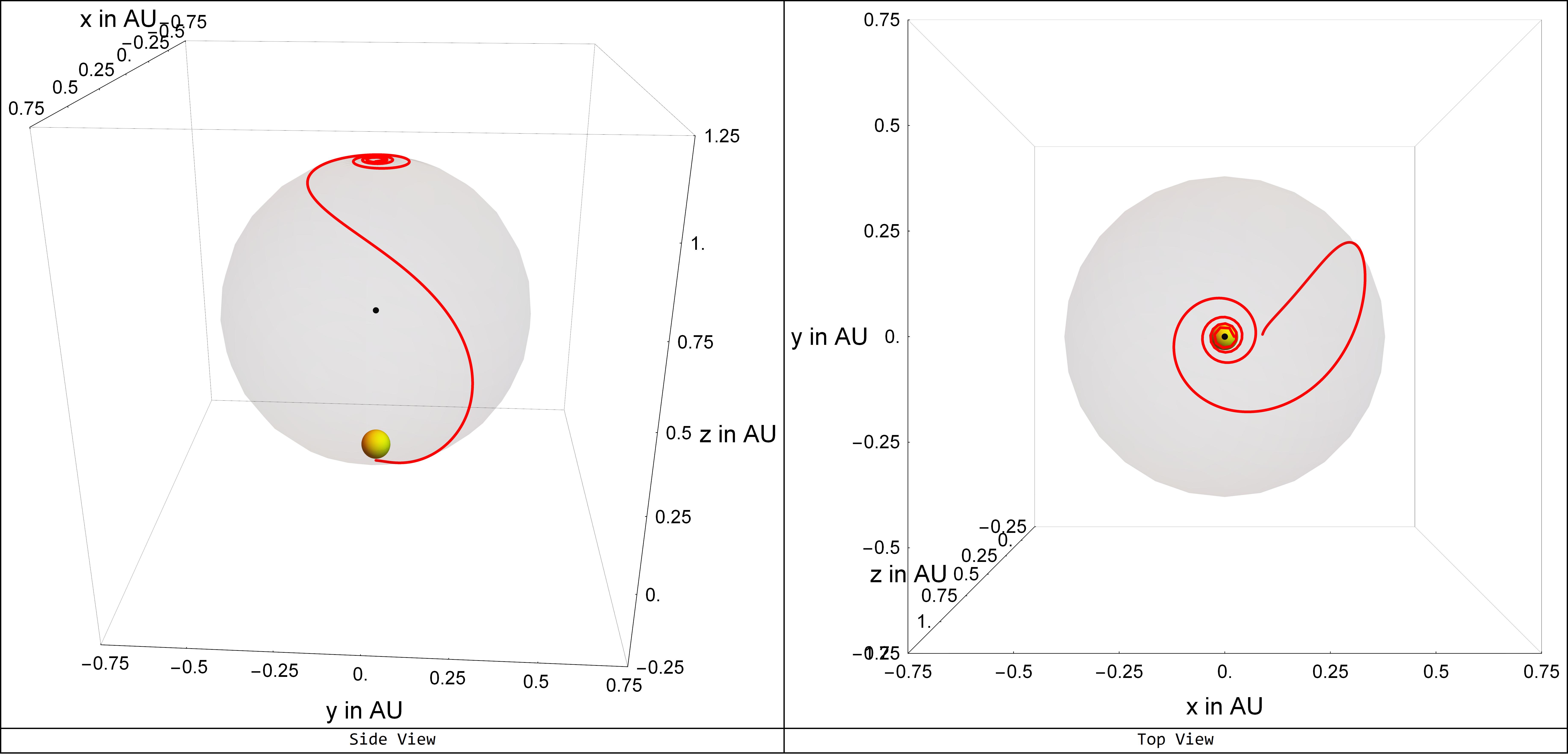}
		\caption{Single trajectory}
		\label{fig:single-traj}
	\end{subfigure}
	\begin{subfigure}{1\textwidth}
		\centering
		\includegraphics[scale=0.32]{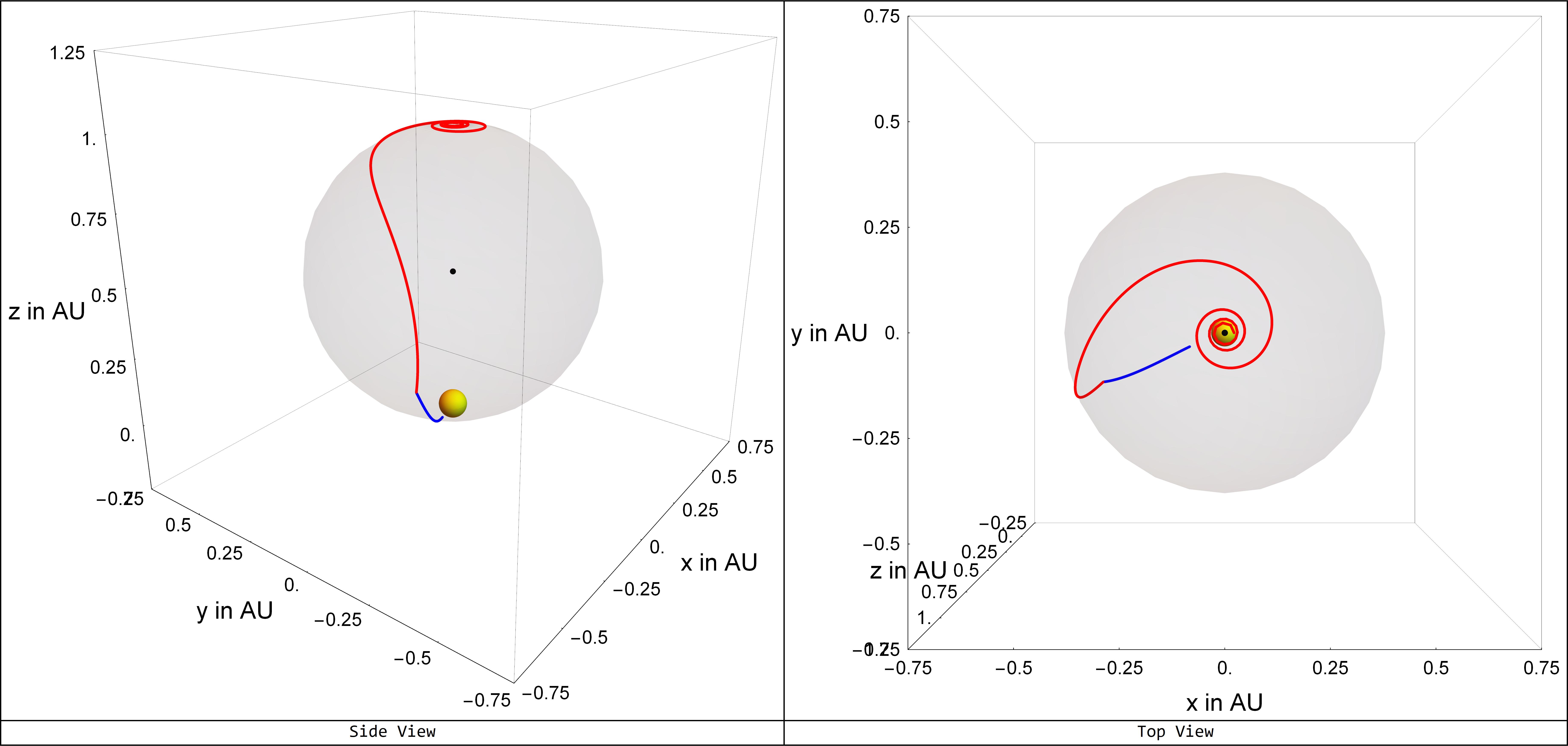}
		\caption{Two-stage trajectory}
		\label{fig:two-stage-traj}
	\end{subfigure}
	\caption{Optimal single and two-stage trajectories where $n_{\theta}$ is periodic with respect to $\theta$.}
	\label{fig:optimal-traj-1-2}
\end{figure*}

\begin{figure*}[h!]
	\centering
	\begin{subfigure}{1\textwidth}
		\centering
		\includegraphics[scale=0.32]{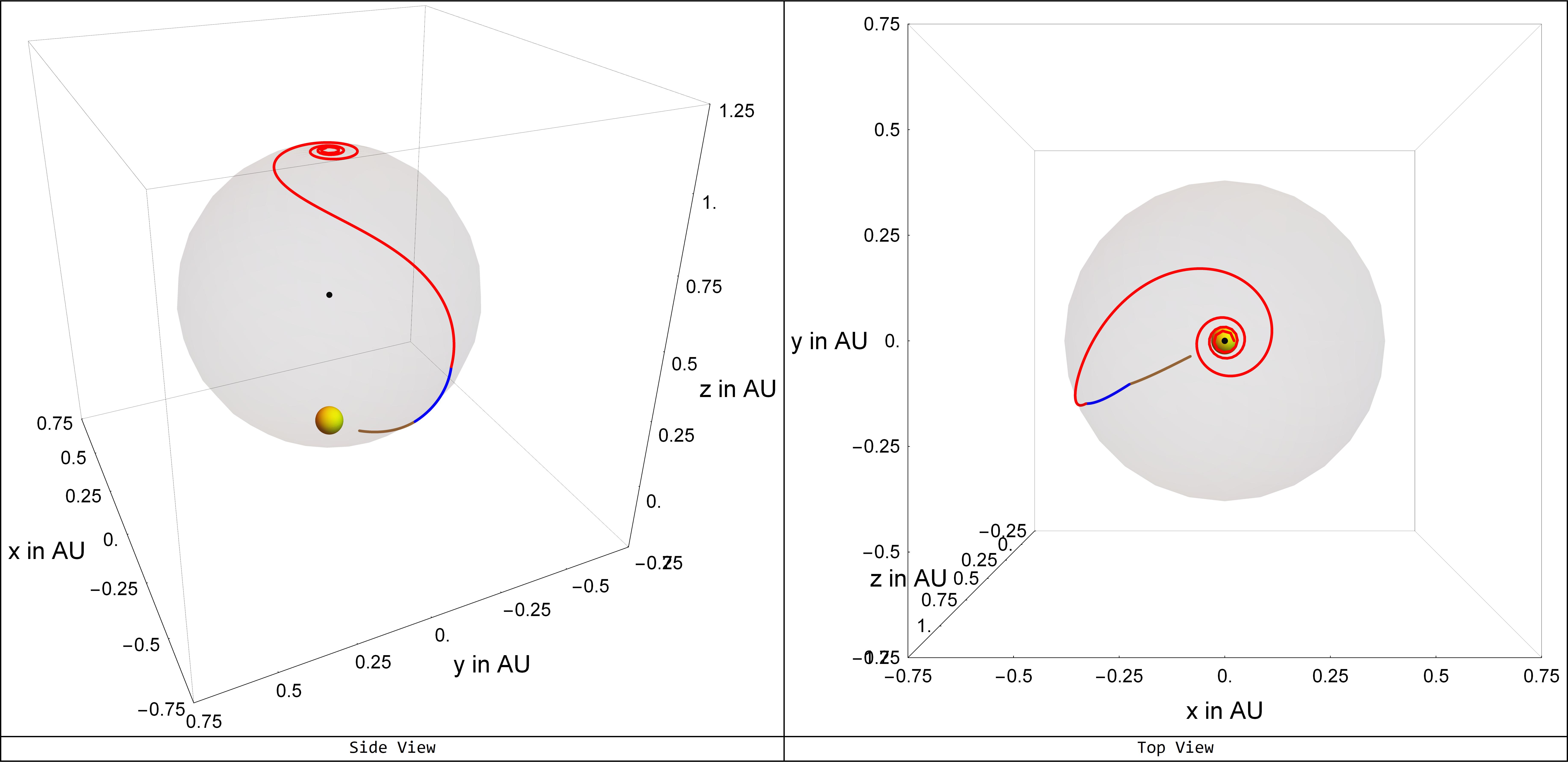}
		\caption{Three-stage trajectory}
		\label{fig:three-stage-traj}
	\end{subfigure}
	\begin{subfigure}{1\textwidth}
		\centering
		\includegraphics[scale=0.32]{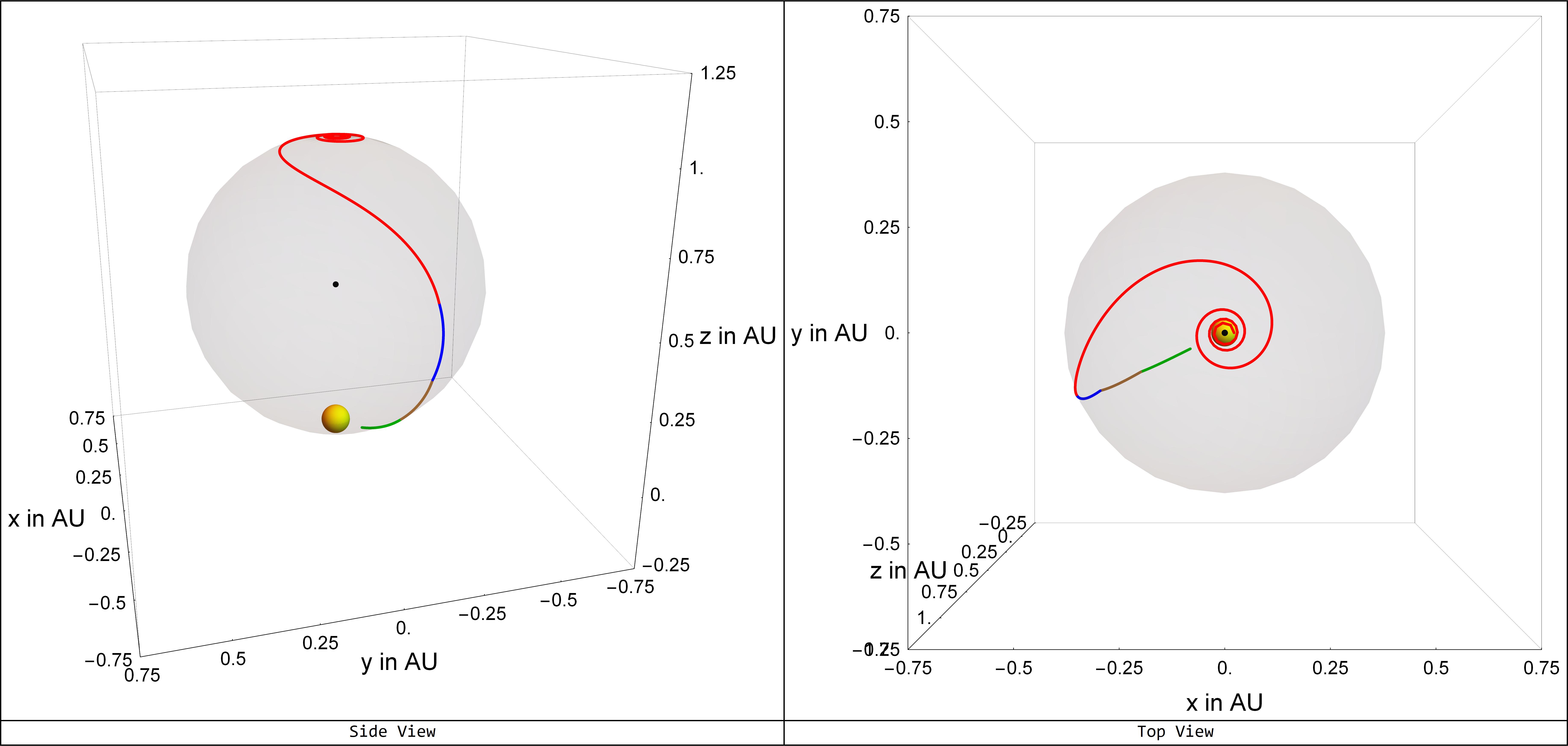}
		\caption{Four-stage trajectory}
		\label{fig:four-stage-traj}
	\end{subfigure}
	\caption{Optimal three-stage and four-stage trajectories where $n_{\theta}$ is periodic with respect to $\theta$.}
	\label{fig:optimal-traj-3-4}
\end{figure*}

The results of the optimization process are shown in Figures \ref{fig:optimal-traj-1-2} and \ref{fig:optimal-traj-3-4}, as well as in Table \ref{tab:solar-probe-mission-deets}. We observe that the flight time decreases upon addition of stages. It should be noted, however, that there is hardly any change in $k$ values from the first two sub-intervals of the four-stage trajectory. Interestingly, this slight change in the $k$ values of the first two parts, making the number of stages four from three, improved the total flight time by a few days. 

\renewcommand\theadfont{}
\begin{table*}[h!]
	\caption{Details for the solar probe mission. The cone angle is set at the optimal cone angle of $-36.25^{\circ}$.}
	\label{tab:solar-probe-mission-deets}
	\centering
	\scalebox{0.95}{
		\begin{tabular}{|l|c|c|c|} 
			\hline
			& \textbf{$k$ Values} & \textbf{Transition Points} & \textbf{Flight Time, d} \\ \hline
			Single		& 1.50968650	& None	& 157.83\\ \hline
			Two-Stage 	& \thead{1.72659908 \\  1.24968798}	& $(0.5217\text{ AU},58.556^{\circ},1282.01^{\circ})$	& 144.34 \\ \hline
			Three-Stage & \thead{1.72659913 \\  1.53541026 \\ 1.20115984}	& \thead{$(0.6480\text{ AU},49.610^{\circ},1284.34^{\circ})$ \\ $(0.4022\text{ AU},66.286^{\circ},1284.49^{\circ})$}	& 142.18 \\ \hline
			Four-Stage 	& \thead{1.72659977 \\ 1.72659709 \\ 1.42118131 \\ 1.18060364}	& \thead{$(0.7559\text{ AU},40.897^{\circ},1282.74^{\circ})$ \\ $(0.5537\text{ AU},56.377^{\circ},1284.94^{\circ})$ \\ $(0.3508\text{ AU},69.463^{\circ},1284.96^{\circ})$}	& 140.95 \\ \hline
		\end{tabular}
	}
\end{table*}

\begin{figure*}[h!]
	\centering
	\begin{subfigure}{0.5\textwidth}
		\centering
		\includegraphics[scale=0.45]{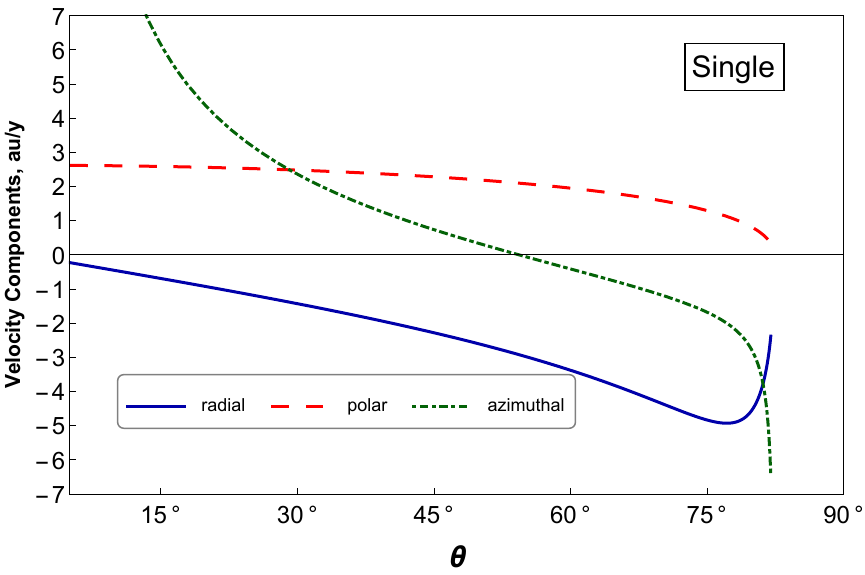}
		\caption{Single stage}
		\label{fig:single-stage-velocity}
	\end{subfigure}%
	\begin{subfigure}{0.5\textwidth}
		\centering
		\includegraphics[scale=0.45]{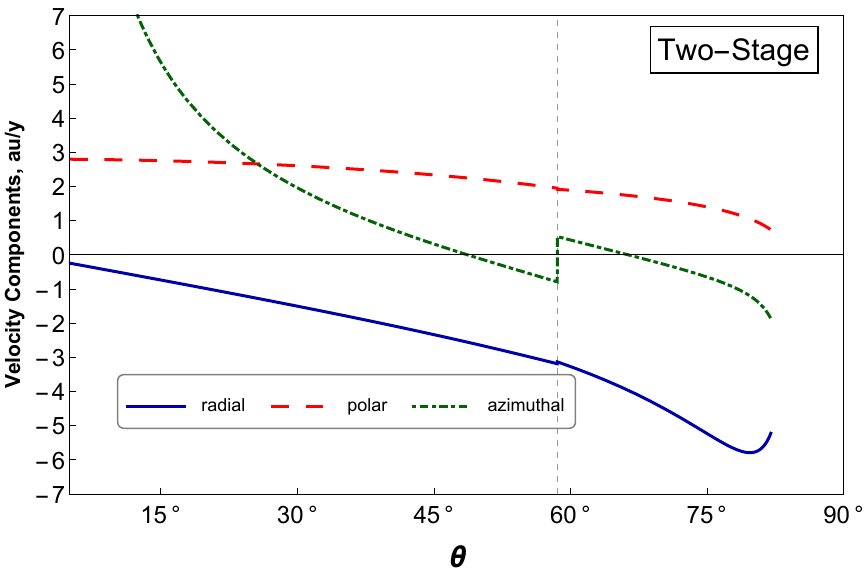}
		\caption{Two-stage}
		\label{fig:two-stage-velocity}
	\end{subfigure}
	\begin{subfigure}{0.5\textwidth}
		\centering
		\includegraphics[scale=0.45]{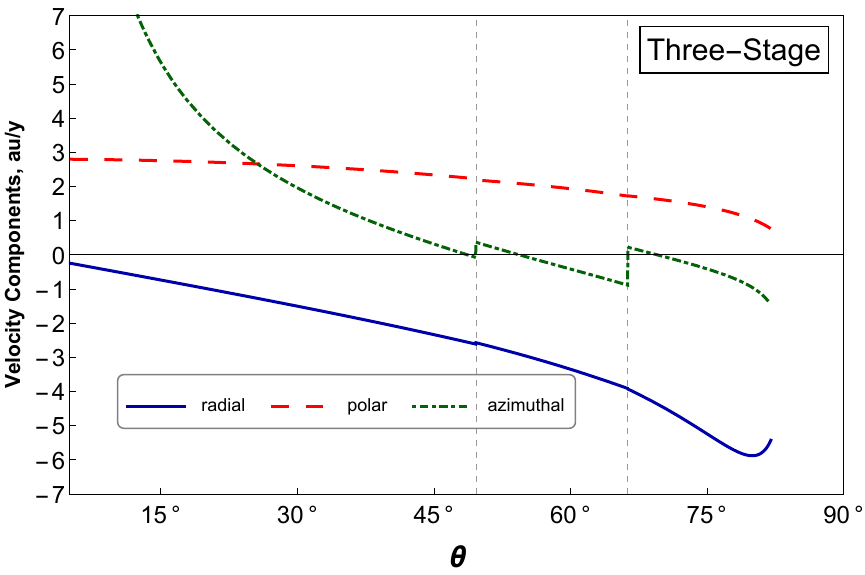}
		\caption{Three-stage}
		\label{fig:three-stage-velocity}
	\end{subfigure}%
	\begin{subfigure}{0.5\textwidth}
		\centering
		\includegraphics[scale=0.45]{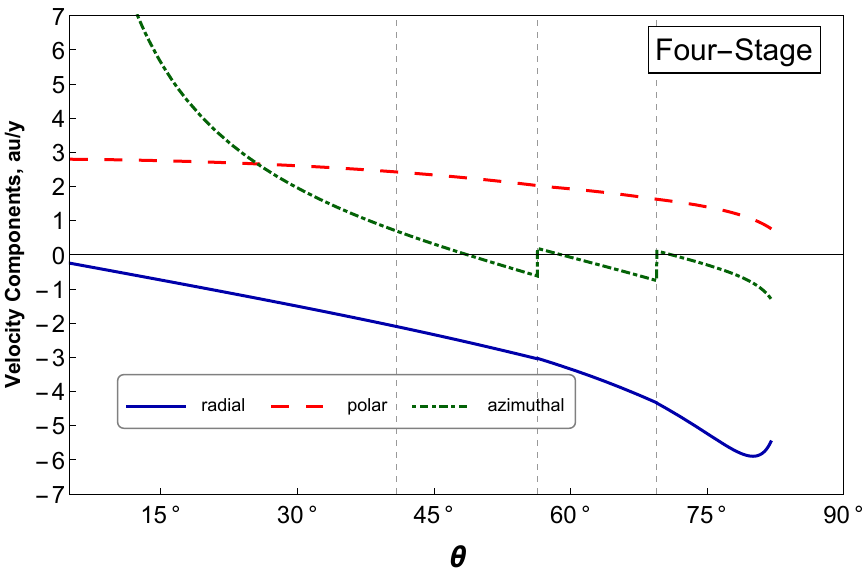}
		\caption{Four-stage}
		\label{fig:four-stage-velocity}
	\end{subfigure}
	\caption{Velocity profiles for an time optimal transfer for a solar sail probe with different number of stages. The vertical dashed lines indicate the transition points.}
	\label{fig:optimal-traj-velocities}
\end{figure*}

We can gain more understanding about these trajectories by looking at the radial, polar, and azimuthal velocities as functions of the polar angle (\textit{See Figure \ref{fig:optimal-traj-velocities}.}) We note that the velocities here are measured with respect to the reference point which is the sun. Obtaining the velocities relative to the earth is straightforward though. The radial velocity is negative, indicating that the sail points towards the direction of the sun, while the polar velocity is positive, consistent with our convention. Because in all of the optimal trajectories, $k>1$, it is expected that the sail reverses its azimuthal direction, indicating a change in sign in the azimuthal velocity plot. The sail starts with a smaller magnitude of $\rdot$ but a large value of azimuthal velocity, implying that at smaller values of $\theta$, the sail's motion is mainly helicoidal. As $\theta$ increases, the azimuthal velocity decreases, making the trajectory follow a more directly downward curve. Consequently, the sail approaches the sun, increasing its speed, consistent with the increase in the magnitude of the radial velocity in the $\rdot$-vs-$\theta$ plot.

\begin{figure*}[h!]
	\centering
	\begin{subfigure}{0.5\textwidth}
		\centering
		\includegraphics[scale=0.45]{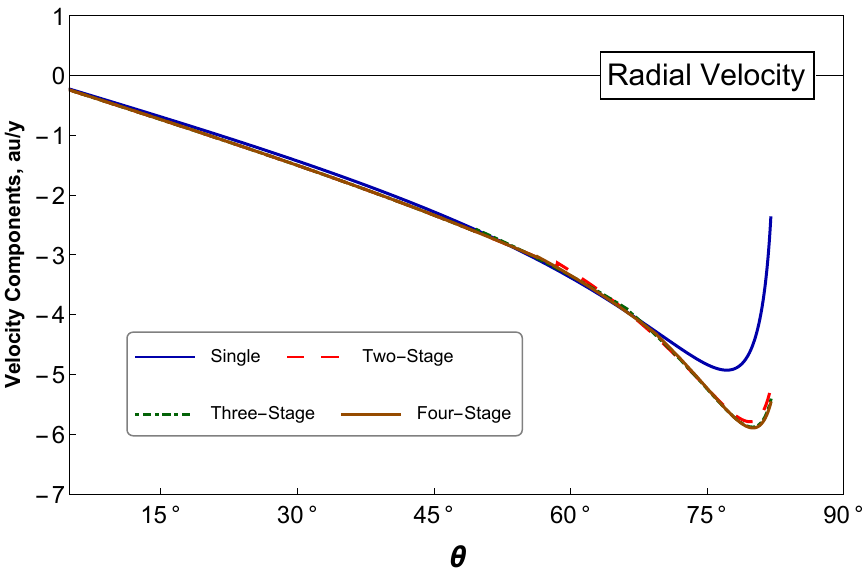}
		\caption{Radial Velocity}
		\label{fig:radial-velocity}
	\end{subfigure}%
	\begin{subfigure}{0.5\textwidth}
		\centering
		\includegraphics[scale=0.45]{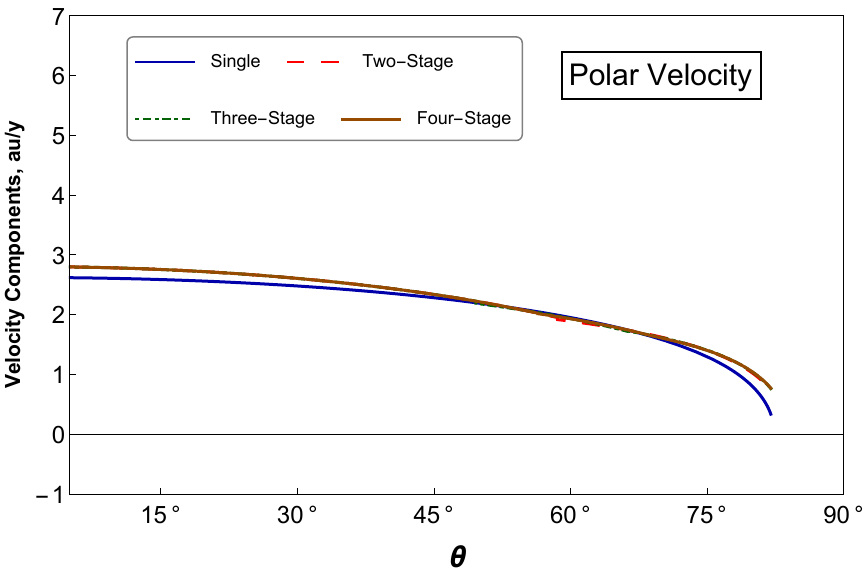}
		\caption{Polar Velocity}
		\label{fig:polar-velocity}
	\end{subfigure}
	\begin{subfigure}{1.0\textwidth}
		\centering
		\includegraphics[scale=0.45]{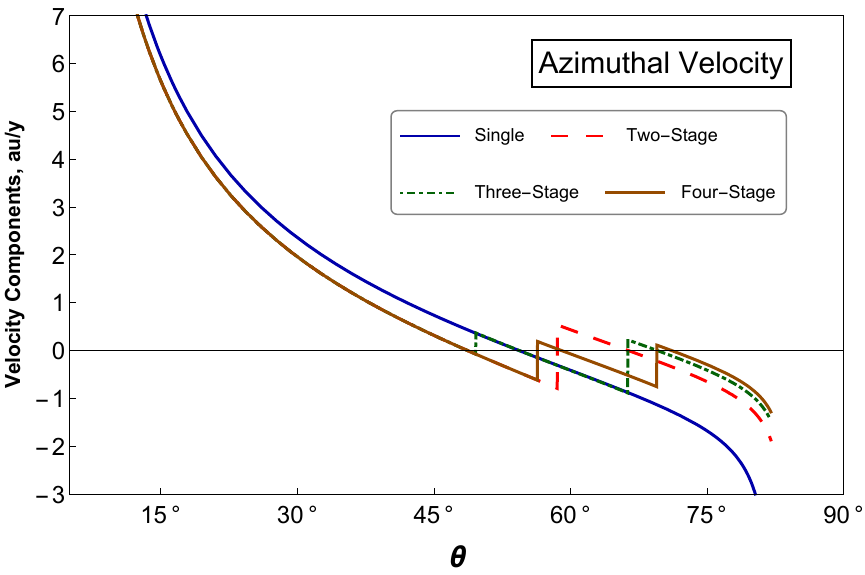}
		\caption{Azimuthal Velocity}
		\label{fig:azimuthal-velocity}
	\end{subfigure}
	\caption{Components of the velocity for different number of stages.}
	\label{fig:optimal-traj-velocities2}
\end{figure*}

We can also compare the velocity components for different number of stages. We re-arranged the plots in Figure \ref{fig:optimal-traj-velocities} to compare the velocity component for different number of stages (\textit{See Figure \ref{fig:optimal-traj-velocities2}}). We observed that while there is a noticeable change in the velocity profile from the single stage to two-stage trajectories, there is a small variation among the velocities in the two-stage, three-stage, and four-stage optimal trajectories. Hence, using a four-stage optimal trajectory gives us the most optimal flight time without expending additional effort in changing the radial and polar valocities of the sail.

A different case is observed in the azimuthal component of the velocity. There is a jump among the different parts of the $r\phidot\sin\theta$-vs-$\theta$ for multiple stage optimal trajectories. These jumps occur at the transition points and are common for the two-stage, three-stage, and four-stage transfers. Hence, for the optimal trajectory to be followed, the sail must drastically change its azimuthal speed near the transition point. 

The transition points are located at $\theta>\pi/4$ where the trajectory is not helicoidal anymore. At these regions, since $k>1$, the sail will reverse its trajectory in the azimuthal direction. By changing its $k$ value at these points, the sail corrects its orientation for it not to immediately reverse its direction, but for it to traverse directly towards the direction of the sun along the spherical surface. This `self-correcting' mechanism makes the total flight time a minimum. 

\section{Conclusions and recommendations}
\label{sec:conc}
In this paper, we used the surface constraint approach in designing and optimizing the trajectories of a solar sailing spacecraft that is constrained on a displaced spherical surface. We showed that assuming that the cone angle of the sail is constant, the generalized Laplace-Runge-Lenz vector is conserved, hence both the radial and azimuthal equations can be determined. By choosing the appropriate constraint equation, we were successful in obtaining a family of trajectories that are highly-dependent on the form of $\delta(\theta)$.

For the case when both the cone and clock angles are constant throughout the mission, we found out that the trajectory becomes more helicoidal as the clock angle increases, which is consistent with the observations in our previous study \citep{garrido2023surface}. This result was also supported by the fact that the azimuthal force increases as the clock angle increases. 

Interesting results were also obtained for the case when the component of the unit normal vector followed equation \eqref{eq:periodic_ntheta_nphi}. For $k<1$, the sail traverses a spiral with increasing number of turns as $k$ decreases. On the other hand, a reversal of the trajectory in the azimuthal direction was observed for $k\ge 1$, which became more pronounced for larger values of $k$. These characteristics were also observed in our previous study on a sail on a cylindrical surface, even if they differ in the form of the constraint equation.

So far, only the constant and periodic components of $n_{\theta}$ have been studied in the literature. However, the surface constraint approach allows for any form of the functional dependence of $\delta$, provided that the normalization condition for the unit normal vector is satisfied. Studying other forms of $n_{\theta}$ can be a topic of future research. By changing the form of $n_{\theta}$, we can get a family of trajectories different from what we currently have. Obtaining these families of trajectories will enrich our library of controls that we can use for optimization. As a consequence, we can formulate an optimization problem in $n_{\theta}$, that is, given same value of the lightness number, the surface constraint equation, and the initial and terminal conditions, what functional dependence of $\delta$ and/or $n_{\theta}$ will give us the minimum flight time?

A common characteristic of the surfaces we considered for the surface constraint approach is that they are all symmetric with respect to the $z$ axis. We can extend our analysis to those surfaces of revolution with the same property. Furthermore, we have not yet considered the case when the constraint equation is also a function of the azimuthal angle $\varphi$. What will happen to the trajectory equations if that is indeed the case?

Finally, we obtained an optimal transfer trajectory for a solar probe mission. Similar to what was done in \citet{garrido2023flight}, we considered split-continuous trajectories, with each stage having a unique $k$ value. We observed that by increasing the number of stages, the flight time becomes smaller. There is a great improvement in the flight time from the single stage to the two-stage optimal trajectories. Interestingly, the number of stages can thus be increased without major changes in the sail's velocity. At the transition points, the sail's azimuthal velocity resets, allowing for the sail to correct its path and approach the sun directly. The velocities we have obtained are comparable to those desired in the literature, for example in \citet{turyshev2023science}. 

The results we have shown here can be used in the preliminary design and optimization of solar sail missions. We hope that our results contribute in planning future missions concerning solar observations. 




\bibliographystyle{jasr-model5-names}
\biboptions{authoryear}
\bibliography{biblio1.bib}

\begin{thebibliography}{50}
\expandafter\ifx\csname natexlab\endcsname\relax\def\natexlab#1{#1}\fi
\ifx\xfnm\relax \def\xfnm[#1]{\unskip,\space#1}\fi

\bibitem[{Alhorn et~al.(2011)Alhorn, Casas, Agasid et~al.}]{alhorn2011nanosail}
\bibinfo{author}{Alhorn, D.}, \bibinfo{author}{Casas, J.},
  \bibinfo{author}{Agasid, E.} et~al. (\bibinfo{year}{2011}).
\newblock \bibinfo{title}{Nanosail-{D}: The small satellite that could!}
\newblock In {\it \bibinfo{booktitle}{Proceedings of the 25th Annual AIAA/USU
  Conference on Small Satellites}\/} Small but Mighty, SSC11-VI-1.
\newblock
  \bibinfo{note}{\url{https://digitalcommons.usu.edu/smallsar/2011/all2012/1/}}.

\bibitem[{Appourchaux et~al.(2009)Appourchaux, Liewer, Watt, Alexander,
  Andretta, Auch{\`e}re, D’Arrigo, Ayon, Corbard, Fineschi
  et~al.}]{appourchaux2009polar}
\bibinfo{author}{Appourchaux, T.}, \bibinfo{author}{Liewer, P.},
  \bibinfo{author}{Watt, M.} et~al. (\bibinfo{year}{2009}).
\newblock \bibinfo{title}{Polar investigation of the sun—polaris}.
\newblock {\it \bibinfo{journal}{Experimental Astronomy}\/},  {\it
  \bibinfo{volume}{23}\/}, \bibinfo{pages}{1079--1117}.
  \DOIprefix\doi{10.1007/s10686-008-9107-8}.

\bibitem[{Baoyin \& McInnes(2006)}]{baoyin2006solar}
\bibinfo{author}{Baoyin, H.},  \& \bibinfo{author}{McInnes, C.~R.}
  (\bibinfo{year}{2006}).
\newblock \bibinfo{title}{Solar sail equilibria in the elliptical restricted
  three-body problem}.
\newblock {\it \bibinfo{journal}{J Guid Control Dynam}\/},  {\it
  \bibinfo{volume}{29}\/}\bibinfo{issue}{(3)}, \bibinfo{pages}{538--543}.
  \DOIprefix\doi{10.2514/1.15596}.

\bibitem[{Betts et~al.(2017)Betts, Nye, Vaughn et~al.}]{betts2017lightsail}
\bibinfo{author}{Betts, B.}, \bibinfo{author}{Nye, B.},
  \bibinfo{author}{Vaughn, J.} et~al. (\bibinfo{year}{2017}).
\newblock \bibinfo{title}{Lightsail 1 mission results and public outreach
  strategies}.
\newblock In {\it \bibinfo{booktitle}{Fourth International Symposium on Solar
  Sailing}\/}.

\bibitem[{Bookless \& McInnes(2008)}]{bookless2008control}
\bibinfo{author}{Bookless, J.},  \& \bibinfo{author}{McInnes, C.}
  (\bibinfo{year}{2008}).
\newblock \bibinfo{title}{Control of {L}agrange point orbits using solar sail
  propulsion}.
\newblock {\it \bibinfo{journal}{Acta Astronaut}\/},  {\it
  \bibinfo{volume}{62}\/}\bibinfo{issue}{(2-3)}, \bibinfo{pages}{159--176}.
  \DOIprefix\doi{10.1016/j.actaastro.2006.12.051}.

\bibitem[{Brent(2013)}]{brent2013algorithms}
\bibinfo{author}{Brent, R.~P.} (\bibinfo{year}{2013}).
\newblock {\it \bibinfo{title}{Algorithms for minimization without
  derivatives}\/}.
\newblock \bibinfo{publisher}{Courier Corporation}.

\bibitem[{Conn et~al.(2000)Conn, Gould \& Toint}]{conn2000trust}
\bibinfo{author}{Conn, A.~R.}, \bibinfo{author}{Gould, N.~I.},  \&
  \bibinfo{author}{Toint, P.~L.} (\bibinfo{year}{2000}).
\newblock {\it \bibinfo{title}{Trust region methods}\/}.
\newblock \bibinfo{publisher}{SIAM}.

\bibitem[{Conway(2010)}]{conway2010spacecraft}
\bibinfo{author}{Conway, B.} (\bibinfo{year}{2010}).
\newblock \bibinfo{title}{Spacecraft trajectory optimization}.
\newblock \bibinfo{publisher}{Cambridge University Press}
  volume~\bibinfo{volume}{29} of {\it \bibinfo{series}{Cambridge Aerospace
  Series}\/}.

\bibitem[{Dachwald(2005)}]{dachwald2005optimal}
\bibinfo{author}{Dachwald, B.} (\bibinfo{year}{2005}).
\newblock \bibinfo{title}{Optimal solar sail trajectories for missions to the
  outer solar system}.
\newblock {\it \bibinfo{journal}{J Guid Control Dynam}\/},  {\it
  \bibinfo{volume}{28}\/}\bibinfo{issue}{(6)}, \bibinfo{pages}{1187--1193}.
  \DOIprefix\doi{10.2514/1.13301}.

\bibitem[{Fan et~al.(2020)Fan, Huo, Qi, Zhao, Yu \& Lin}]{fan2020initial}
\bibinfo{author}{Fan, Z.}, \bibinfo{author}{Huo, M.}, \bibinfo{author}{Qi, N.}
  et~al. (\bibinfo{year}{2020}).
\newblock \bibinfo{title}{Initial design of low-thrust trajectories based on
  the bezier curve-based shaping approach}.
\newblock {\it \bibinfo{journal}{Proceedings of the Institution of Mechanical
  Engineers, Part G: J Aerospace Eng}\/},  {\it
  \bibinfo{volume}{234}\/}\bibinfo{issue}{(11)}, \bibinfo{pages}{1825--1835}.
  \DOIprefix\doi{10.1177/0954410020920040}.

\bibitem[{Forsythe et~al.(1977)}]{forsythe1977computer}
\bibinfo{author}{Forsythe, G.~E.} et~al. (\bibinfo{year}{1977}).
\newblock {\it \bibinfo{title}{Computer methods for mathematical
  computations}\/}.
\newblock \bibinfo{publisher}{Prentice-hall}.

\bibitem[{Fu et~al.(2016)Fu, Sperber \& Eke}]{fu2016solar}
\bibinfo{author}{Fu, B.}, \bibinfo{author}{Sperber, E.},  \&
  \bibinfo{author}{Eke, F.} (\bibinfo{year}{2016}).
\newblock \bibinfo{title}{Solar sail technology-a state of the art review}.
\newblock {\it \bibinfo{journal}{Prog Aerosp Sci}\/},  {\it
  \bibinfo{volume}{86}\/}, \bibinfo{pages}{1--19}.
  \DOIprefix\doi{10.1016/j.paerosci.2016.07.001}.

\bibitem[{Garrido \& Esguerra(2023{\natexlab{a}})}]{garrido2023surface}
\bibinfo{author}{Garrido, J.},  \& \bibinfo{author}{Esguerra, J.~P.}
  (\bibinfo{year}{2023}{\natexlab{a}}).
\newblock \bibinfo{title}{A surface constraint approach for solar sail
  trajectories}.
\newblock {\it \bibinfo{journal}{Advances in Space Research}\/},  {\it
  \bibinfo{volume}{71}\/}\bibinfo{issue}{(10)}, \bibinfo{pages}{4256--4275}.
  \DOIprefix\doi{10.1016/j.asr.2022.12.041}.

\bibitem[{Garrido \& Esguerra(2023{\natexlab{b}})}]{garrido2023flight}
\bibinfo{author}{Garrido, J.~V.},  \& \bibinfo{author}{Esguerra, J. P.~H.}
  (\bibinfo{year}{2023}{\natexlab{b}}).
\newblock \bibinfo{title}{Flight time optimization of cylindrical constrained
  solar sail trajectories}.
\newblock In {\it \bibinfo{booktitle}{Proceedings of the Samahang Pisika ng
  Pilipinas}\/} (pp. \bibinfo{pages}{SPP--2023--PB--26}).
\newblock volume~\bibinfo{volume}{41}.
\newblock \URLprefix
  \url{https://proceedings.spp-online.org/article/view/SPP-2023-PB-26}.

\bibitem[{Gong \& Macdonald(2019)}]{gong2019review}
\bibinfo{author}{Gong, S.},  \& \bibinfo{author}{Macdonald, M.}
  (\bibinfo{year}{2019}).
\newblock \bibinfo{title}{Review on solar sail technology}.
\newblock {\it \bibinfo{journal}{Astrodynam}\/},  {\it \bibinfo{volume}{3}\/},
  \bibinfo{pages}{93--125}. \DOIprefix\doi{10.1007/s42064-019-0038-x}.

\bibitem[{Gong et~al.(2011)Gong, Li \& Zeng}]{gong2011utilization}
\bibinfo{author}{Gong, S.-P.}, \bibinfo{author}{Li, J.-F.},  \&
  \bibinfo{author}{Zeng, X.-Y.} (\bibinfo{year}{2011}).
\newblock \bibinfo{title}{Utilization of an {H}-reversal trajectory of a solar
  sail for asteroid deflection}.
\newblock {\it \bibinfo{journal}{Res Astron Astrophys}\/},  {\it
  \bibinfo{volume}{11}\/}\bibinfo{issue}{(10)}, \bibinfo{pages}{1123}.
  \DOIprefix\doi{10.1088/1674-4527/11/10/001}.

\bibitem[{Heiligers et~al.(2019)Heiligers, Fernandez, Stohlman \&
  Wilkie}]{heiligers2019trajectory}
\bibinfo{author}{Heiligers, J.}, \bibinfo{author}{Fernandez, J.~M.},
  \bibinfo{author}{Stohlman, O.~R.} et~al. (\bibinfo{year}{2019}).
\newblock \bibinfo{title}{Trajectory design for a solar-sail mission to
  asteroid 2016 {HO3}}.
\newblock {\it \bibinfo{journal}{Astrodynam}\/},  {\it
  \bibinfo{volume}{3}\/}\bibinfo{issue}{(3)}, \bibinfo{pages}{231--246}.

\bibitem[{Izzo(2006)}]{izzo2006lambert}
\bibinfo{author}{Izzo, D.} (\bibinfo{year}{2006}).
\newblock \bibinfo{title}{Lambert's problem for exponential sinusoids}.
\newblock {\it \bibinfo{journal}{J Guid Control Dynam}\/},  {\it
  \bibinfo{volume}{29}\/}, \bibinfo{pages}{1242--1245}.
  \DOIprefix\doi{10.2514/1.21796}.

\bibitem[{Kobayashi et~al.(2020)Kobayashi, Johnson, Thomas, McIntosh, McKenzie,
  Newmark, Wright~Jr, Bean, Fabisinski, Capizzo et~al.}]{kobayashi2020high}
\bibinfo{author}{Kobayashi, K.}, \bibinfo{author}{Johnson, L.},
  \bibinfo{author}{Thomas, H.~D.} et~al. (\bibinfo{year}{2020}).
\newblock \bibinfo{title}{The high inclination solar mission (hism)}.
\newblock In {\it \bibinfo{booktitle}{AGU Fall Meeting Abstracts}\/} (pp.
  \bibinfo{pages}{SH011--0004}).
\newblock volume \bibinfo{volume}{2020}.

\bibitem[{Lappas et~al.(2009)Lappas, Mengali, Quarta
  et~al.}]{lappas2009practical}
\bibinfo{author}{Lappas, V.}, \bibinfo{author}{Mengali, G.},
  \bibinfo{author}{Quarta, A.} et~al. (\bibinfo{year}{2009}).
\newblock \bibinfo{title}{Practical systems design for an
  {E}arth-magnetotail-monitoring solar sail mission}.
\newblock {\it \bibinfo{journal}{J Spacecraft Rockets}\/},  {\it
  \bibinfo{volume}{46}\/}, \bibinfo{pages}{381--399}.
  \DOIprefix\doi{10.2514/1.32040}.

\bibitem[{Liewer et~al.(2013)Liewer, Alexander, Appourchaux, Ayon, Floyd,
  Hassler, Kosovichev, Lugaz, Leibacher, Murphy et~al.}]{liewer2013solar}
\bibinfo{author}{Liewer, P.}, \bibinfo{author}{Alexander, D.},
  \bibinfo{author}{Appourchaux, T.} et~al. (\bibinfo{year}{2013}).
\newblock \bibinfo{title}{A solar polar imager concept: Observing solar
  activity from a new perspective}.
\newblock {\it \bibinfo{journal}{NASA, Tech. Rep}\/}, .

\bibitem[{Lingam \& Loeb(2020)}]{lingam2020propulsion}
\bibinfo{author}{Lingam, M.},  \& \bibinfo{author}{Loeb, A.}
  (\bibinfo{year}{2020}).
\newblock \bibinfo{title}{Propulsion of spacecraft to relativistic speeds using
  natural astrophysical sources}.
\newblock {\it \bibinfo{journal}{Astrophys J}\/},  {\it
  \bibinfo{volume}{894}\/}\bibinfo{issue}{(1)}, \bibinfo{pages}{36}.
  \DOIprefix\doi{10.3847/1538-4357/ab7dc7}.

\bibitem[{Macdonald et~al.(2007)Macdonald, Hughes, Falkner \&
  Atzei}]{macdonald2007geosail}
\bibinfo{author}{Macdonald, M.}, \bibinfo{author}{Hughes, G.},
  \bibinfo{author}{Falkner, P.} et~al. (\bibinfo{year}{2007}).
\newblock \bibinfo{title}{Geo{S}ail: an elegant solar sail demonstration
  mission}.
\newblock {\it \bibinfo{journal}{J Spacecraft Rockets}\/},  {\it
  \bibinfo{volume}{44}\/}, \bibinfo{pages}{784--796}.
  \DOIprefix\doi{10.2514/1.22867}.

\bibitem[{Macdonald et~al.(2010)Macdonald, McInnes \&
  Hughes}]{macdonald2010technology}
\bibinfo{author}{Macdonald, M.}, \bibinfo{author}{McInnes, C.},  \&
  \bibinfo{author}{Hughes, G.} (\bibinfo{year}{2010}).
\newblock \bibinfo{title}{Technology requirements of exploration beyond
  {N}eptune by solar sail propulsion}.
\newblock {\it \bibinfo{journal}{J Spacecraft Rockets}\/},  {\it
  \bibinfo{volume}{47}\/}\bibinfo{issue}{(3)}, \bibinfo{pages}{472--483}.
  \DOIprefix\doi{10.2514/1.46657}.

\bibitem[{McInnes(1998)}]{mcinnes1998dynamics}
\bibinfo{author}{McInnes, C.} (\bibinfo{year}{1998}).
\newblock \bibinfo{title}{Dynamics, stability, and control of displaced
  non-{K}eplerian orbits}.
\newblock {\it \bibinfo{journal}{J Guid Control Dynam}\/},  {\it
  \bibinfo{volume}{21}\/}, \bibinfo{pages}{799--805}.
  \DOIprefix\doi{10.2514/2.4309}.

\bibitem[{McInnes(2004)}]{mcinnes2004solar}
\bibinfo{author}{McInnes, C.} (\bibinfo{year}{2004}).
\newblock {\it \bibinfo{title}{Solar sailing: technology, dynamics and mission
  applications}\/}.
\newblock \bibinfo{publisher}{Springer Science \& Business Media}.

\bibitem[{McInnes \& Simmons(1992)}]{mcinnes1992solar}
\bibinfo{author}{McInnes, C.~R.},  \& \bibinfo{author}{Simmons, J.~F.}
  (\bibinfo{year}{1992}).
\newblock \bibinfo{title}{Solar sail halo orbits part {II}-{G}eocentric case}.
\newblock {\it \bibinfo{journal}{J Spacecraft Rockets}\/},  {\it
  \bibinfo{volume}{29}\/}\bibinfo{issue}{(4)}, \bibinfo{pages}{472--479}.
  \DOIprefix\doi{10.2514/3.55639}.

\bibitem[{McKay et~al.(2011)McKay, Macdonald, Biggs et~al.}]{mckay2011survey}
\bibinfo{author}{McKay, R.}, \bibinfo{author}{Macdonald, M.},
  \bibinfo{author}{Biggs, J.} et~al. (\bibinfo{year}{2011}).
\newblock \bibinfo{title}{Survey of highly non-{K}eplerian orbits with
  low-thrust propulsion}.
\newblock {\it \bibinfo{journal}{J Guid Control Dynam}\/},  {\it
  \bibinfo{volume}{34}\/}, \bibinfo{pages}{645--666}.
  \DOIprefix\doi{10.2514/1.52133}.

\bibitem[{Mori et~al.(2020)Mori, Matsumoto, Chujo, Matsushita, Kato, Saiki,
  Tsuda, Kawaguchi, Terui, Mimasu et~al.}]{mori2020solar}
\bibinfo{author}{Mori, O.}, \bibinfo{author}{Matsumoto, J.},
  \bibinfo{author}{Chujo, T.} et~al. (\bibinfo{year}{2020}).
\newblock \bibinfo{title}{Solar power sail mission of {OKEANOS}}.
\newblock {\it \bibinfo{journal}{Astrodynam}\/},  {\it
  \bibinfo{volume}{4}\/}\bibinfo{issue}{(3)}, \bibinfo{pages}{233--248}.
  \DOIprefix\doi{doi.org/10.1007/s42064-019-0067-8}.

\bibitem[{Morrow et~al.(2001)Morrow, Acheeres \& Lubin}]{morrow2001solar}
\bibinfo{author}{Morrow, E.}, \bibinfo{author}{Acheeres, D.},  \&
  \bibinfo{author}{Lubin, D.} (\bibinfo{year}{2001}).
\newblock \bibinfo{title}{Solar sail orbit as asteroids}.
\newblock {\it \bibinfo{journal}{J Spacecraft Rockets}\/},  {\it
  \bibinfo{volume}{38}\/}, \bibinfo{pages}{279--286}.
  \DOIprefix\doi{10.2514/2.3682}.

\bibitem[{Ozimek et~al.(2009)Ozimek, Grebow \& Howell}]{ozimek2009design}
\bibinfo{author}{Ozimek, M.}, \bibinfo{author}{Grebow, D.},  \&
  \bibinfo{author}{Howell, K.} (\bibinfo{year}{2009}).
\newblock \bibinfo{title}{Design of solar sail trajectories with applications
  to {L}unar {S}outh {P}ole convergence}.
\newblock {\it \bibinfo{journal}{J Guid Control Dynam}\/},  {\it
  \bibinfo{volume}{32}\/}, \bibinfo{pages}{1884--1897}.
  \DOIprefix\doi{10.2514/1.41963}.

\bibitem[{Peloni et~al.(2016)Peloni, Ceriotti \& Dachwald}]{peloni2016solar}
\bibinfo{author}{Peloni, A.}, \bibinfo{author}{Ceriotti, M.},  \&
  \bibinfo{author}{Dachwald, B.} (\bibinfo{year}{2016}).
\newblock \bibinfo{title}{Solar-sail trajectory design for a multiple
  near-earth-asteroid rendezvous mission}.
\newblock {\it \bibinfo{journal}{J Guid Control Dynam}\/},  {\it
  \bibinfo{volume}{39}\/}, \bibinfo{pages}{2712--2724}.
  \DOIprefix\doi{10.2514/1.G000470}.

\bibitem[{Petropoulos(2003)}]{petropoulos2003shape}
\bibinfo{author}{Petropoulos, A.} (\bibinfo{year}{2003}).
\newblock {\it \bibinfo{title}{A shape-based approach to automated, low-thrust,
  gravity-assist trajectory design.}\/}.
\newblock Ph.D. thesis School of Aeronautics and Astronautics, Purdue
  University \bibinfo{address}{West Lafayette, IN}.

\bibitem[{Petropoulos \& Longuski(2004)}]{petropoulos2004shape}
\bibinfo{author}{Petropoulos, A.},  \& \bibinfo{author}{Longuski, J.}
  (\bibinfo{year}{2004}).
\newblock \bibinfo{title}{Shape-based algorithm for the automated design of
  low-thrust, gravity assist trajectories}.
\newblock {\it \bibinfo{journal}{J. Spacecraft Rockets}\/},  {\it
  \bibinfo{volume}{41}\/}, \bibinfo{pages}{787--796}.
  \DOIprefix\doi{10.2514/1.13095}.

\bibitem[{Petropoulos \& Sims(2002)}]{petropoulos2002review}
\bibinfo{author}{Petropoulos, A.},  \& \bibinfo{author}{Sims, J.}
  (\bibinfo{year}{2002}).
\newblock \bibinfo{title}{A review of some exact solutions to the planar
  equations of motion of a thrusting spacecraft}.
\newblock In {\it \bibinfo{booktitle}{Proceedings of the the 2nd International
  Symposium on Low Thrust Trajectories}\/}.
\newblock \bibinfo{address}{Toulouse, France}.

\bibitem[{Pezent et~al.(2018)Pezent, Sood \& Heaton}]{pezent2018near}
\bibinfo{author}{Pezent, J.~B.}, \bibinfo{author}{Sood, R.},  \&
  \bibinfo{author}{Heaton, A.} (\bibinfo{year}{2018}).
\newblock \bibinfo{title}{Near {E}arth {A}steroid ({NEA}) {S}cout solar sail
  contingency trajectory design and analysis}.
\newblock In {\it \bibinfo{booktitle}{2018 Space Flight Mechanics Meeting}\/}
  (p. \bibinfo{pages}{0199}).
\newblock \DOIprefix\doi{10.2514/6.2018-0199}.

\bibitem[{Song \& Gong(2019)}]{song2019solar}
\bibinfo{author}{Song, Y.},  \& \bibinfo{author}{Gong, S.}
  (\bibinfo{year}{2019}).
\newblock \bibinfo{title}{Solar-sail trajectory design for multiple near-earth
  asteroid exploration based on deep neural networks}.
\newblock {\it \bibinfo{journal}{Aerosp Sci Technol}\/},  {\it
  \bibinfo{volume}{91}\/}, \bibinfo{pages}{28--40}.
  \DOIprefix\doi{10.1016/j.ast.2019.04.056}.

\bibitem[{Spencer et~al.(2021)Spencer, Betts, Bellardo, Diaz, Plante \&
  Mansell}]{spencer2021lightsail}
\bibinfo{author}{Spencer, D.~A.}, \bibinfo{author}{Betts, B.},
  \bibinfo{author}{Bellardo, J.~M.} et~al. (\bibinfo{year}{2021}).
\newblock \bibinfo{title}{The {L}ightsail 2 solar sailing technology
  demonstration}.
\newblock {\it \bibinfo{journal}{Adv Space Res}\/},  {\it
  \bibinfo{volume}{67}\/}\bibinfo{issue}{(9)}, \bibinfo{pages}{2878--2889}.
  \DOIprefix\doi{10.1016/j.asr.2020.06.029}.

\bibitem[{Taheri \& Abdelkhalik(2012)}]{taheri2012shape}
\bibinfo{author}{Taheri, E.},  \& \bibinfo{author}{Abdelkhalik, O.}
  (\bibinfo{year}{2012}).
\newblock \bibinfo{title}{Shape based approximation of constrained low-thrust
  space trajectories using {F}ourier series}.
\newblock {\it \bibinfo{journal}{J Spacecraft Rockets}\/},  {\it
  \bibinfo{volume}{49}\/}, \bibinfo{pages}{535--546}.
  \DOIprefix\doi{10.2514/1.58789}.

\bibitem[{Thomas et~al.(2020)Thomas, Kobayashi, Mike, Bean, Capizzo, Clements,
  Fabisinski, Garcia \& Steve}]{thomas2020solar}
\bibinfo{author}{Thomas, D.}, \bibinfo{author}{Kobayashi, K.},
  \bibinfo{author}{Mike, B.} et~al. (\bibinfo{year}{2020}).
\newblock \bibinfo{title}{Solar polar imager concept}.
\newblock In {\it \bibinfo{booktitle}{ASCEND 2020}\/} (p.
  \bibinfo{pages}{4060}).
\newblock \DOIprefix\doi{10.2514/6.2020-4060}.

\bibitem[{Tsuda et~al.(2011)Tsuda, Mori, Funase et~al.}]{tsuda2011flight}
\bibinfo{author}{Tsuda, Y.}, \bibinfo{author}{Mori, O.},
  \bibinfo{author}{Funase, R.} et~al. (\bibinfo{year}{2011}).
\newblock \bibinfo{title}{Flight status of {IKAROS} deep space solar sail
  demonstrator}.
\newblock {\it \bibinfo{journal}{Acta Astronaut}\/},  {\it
  \bibinfo{volume}{69}\/}, \bibinfo{pages}{833--840}.
  \DOIprefix\doi{10.1016/j.actaastro.2011.06.005}.

\bibitem[{Turyshev et~al.(2023)Turyshev, Garber, Friedman, Hein, Barnes,
  Batygin, Brown, Cronin, Davoyan, Dubill et~al.}]{turyshev2023science}
\bibinfo{author}{Turyshev, S.~G.}, \bibinfo{author}{Garber, D.},
  \bibinfo{author}{Friedman, L.~D.} et~al. (\bibinfo{year}{2023}).
\newblock \bibinfo{title}{Science opportunities with solar sailing smallsats}.
\newblock {\it \bibinfo{journal}{Planetary and Space Science}\/},  (p.
  \bibinfo{pages}{105744}). \DOIprefix\doi{10.1016/j.pss.2023.105744}.

\bibitem[{Vasile et~al.(2007)Vasile, De~Pascale \&
  Casotto}]{vasile2007optimality}
\bibinfo{author}{Vasile, M.}, \bibinfo{author}{De~Pascale, P.},  \&
  \bibinfo{author}{Casotto, S.} (\bibinfo{year}{2007}).
\newblock \bibinfo{title}{On the optimality of a shape-based approach based on
  pseudo-equinoctial elements}.
\newblock {\it \bibinfo{journal}{Acta Astronaut}\/},  {\it
  \bibinfo{volume}{61}\/}, \bibinfo{pages}{286--297}.
  \DOIprefix\doi{10.1016/j.actaastro.2007.01.017}.

\bibitem[{Vulpetti(2012)}]{vulpetti2012fast}
\bibinfo{author}{Vulpetti, G.} (\bibinfo{year}{2012}).
\newblock \bibinfo{title}{Fast solar sailing: astrodynamics of special
  sailcraft trajectories}.
\newblock \bibinfo{publisher}{Springer Science \& Business Media}
  volume~\bibinfo{volume}{30} of {\it \bibinfo{series}{Space Technology
  Library}\/}.

\bibitem[{Vulpetti et~al.(2014)Vulpetti, Johnson \&
  Matloff}]{vulpetti2014solar}
\bibinfo{author}{Vulpetti, G.}, \bibinfo{author}{Johnson, L.},  \&
  \bibinfo{author}{Matloff, G.} (\bibinfo{year}{2014}).
\newblock {\it \bibinfo{title}{Solar sails: a novel approach to interplanetary
  travel}\/}.
\newblock \bibinfo{publisher}{Springer}.

\bibitem[{Waters \& McInnes(2007)}]{waters2007periodic}
\bibinfo{author}{Waters, T.~J.},  \& \bibinfo{author}{McInnes, C.~R.}
  (\bibinfo{year}{2007}).
\newblock \bibinfo{title}{Periodic orbits above the ecliptic in the solar-sail
  restricted three-body problem}.
\newblock {\it \bibinfo{journal}{J Guid Control Dynam}\/},  {\it
  \bibinfo{volume}{30}\/}\bibinfo{issue}{(3)}, \bibinfo{pages}{687--693}.
  \DOIprefix\doi{10.2514/1.26232}.

\bibitem[{West(2004)}]{west2004geostorm}
\bibinfo{author}{West, J.} (\bibinfo{year}{2004}).
\newblock \bibinfo{title}{The {G}eostorm {W}arning {M}ission: enhanced
  opportunities on new technology}.
\newblock In {\it \bibinfo{booktitle}{14th AAS/AIAA Space Flight Mechanics
  Conference}\/} (pp. \bibinfo{pages}{1--14}).

\bibitem[{Zeng et~al.(2016)Zeng, Gong, Li \& Alfriend}]{zeng2016solar}
\bibinfo{author}{Zeng, X.}, \bibinfo{author}{Gong, S.}, \bibinfo{author}{Li,
  J.} et~al. (\bibinfo{year}{2016}).
\newblock \bibinfo{title}{Solar sail body-fixed hovering over elongated
  asteroids}.
\newblock {\it \bibinfo{journal}{J Guid Control Dynam}\/},  {\it
  \bibinfo{volume}{39}\/}\bibinfo{issue}{(6)}, \bibinfo{pages}{1223--1231}.
  \DOIprefix\doi{10.2514/1.G001061}.

\bibitem[{Zeng et~al.(2019)Zeng, Vulpetti \& Circi}]{zeng2019solar}
\bibinfo{author}{Zeng, X.}, \bibinfo{author}{Vulpetti, G.},  \&
  \bibinfo{author}{Circi, C.} (\bibinfo{year}{2019}).
\newblock \bibinfo{title}{Solar sail {H}-reversal trajectory: A review of its
  advances and applications}.
\newblock {\it \bibinfo{journal}{Astrodynam}\/},  {\it
  \bibinfo{volume}{3}\/}\bibinfo{issue}{(1)}, \bibinfo{pages}{1--15}.
  \DOIprefix\doi{10.1007/s42064-018-0032-y}.

\bibitem[{Zeng et~al.(2015)Zeng, Jiang \& Li}]{zeng2015asteroid}
\bibinfo{author}{Zeng, X.-Y.}, \bibinfo{author}{Jiang, F.-H.},  \&
  \bibinfo{author}{Li, J.-F.} (\bibinfo{year}{2015}).
\newblock \bibinfo{title}{Asteroid body-fixed hovering using nonideal solar
  sails}.
\newblock {\it \bibinfo{journal}{Res Astron Astrophys}\/},  {\it
  \bibinfo{volume}{15}\/}\bibinfo{issue}{(4)}, \bibinfo{pages}{597}.
  \DOIprefix\doi{10.1088/1674-4527/15/4/011}.

\end{thebibliography}

\end{document}